\documentclass{aa}
\usepackage{url}
\usepackage[colorlinks,citecolor=blue]{hyperref}
\usepackage{graphicx}
\usepackage[dvipsnames]{xcolor}
\usepackage{amsmath}
\usepackage{natbib}
\usepackage{txfonts}
\usepackage{float}
\usepackage{adjustbox}
\usepackage{MnSymbol} 
\usepackage{mhchem} 
\usepackage{physics}
\usepackage{placeins} 

\begin{document}

\title{The two-dimensional pressure structure of the HD 163296 protoplanetary disk as~probed by multi-molecule kinematics} 
\author{V. Pezzotta \inst{1}
  \and S. Facchini \inst{1} 
  \and C. Longarini \inst{2} 
    \and G. Lodato \inst{1}
    \and P. Martire \inst{3}}
\institute{Dipartimento di Fisica, Università degli Studi di Milano, Via Celoria 16, Milano, 20133, Italy\\
              \email{viviana.pezzotta@unimi.it}
 \and {Institute of Astronomy, University of Cambridge, Madingley Road, Cambridge, CB3 0HA, United Kingdom}
 \and {Leiden Observatory, Leiden University, Einsteinweg 55, 2333 CC Leiden, The Netherlands}}
\date{Received x xxxxxx xxxx/ Accepted x xxxxxxx xxxx}
\abstract {Gas kinematics is a new and unique way to study planet-forming environments by an accurate characterization of disk velocity fields. High angular resolution ALMA observations allow deep kinematical analysis of disks, by observing molecular line emission at high spectral resolution. In particular, rotation curves are key tools for studying the disk pressure structure and efficiently estimating fundamental disk parameters, such as mass and radius.
In this work we explore the potential of a multi-molecule approach to gas kinematics to provide a 2D characterization of the HD 163296 disk. From the high quality data of the MAPS Large Program we extracted the rotation curves of rotational lines from seven distinct molecular species (\ce{^12}CO, \ce{^13}CO, \ce{C^18O}, HCN, \ce{H2CO}, \ce{HCO+}, \ce{C2H}), spanning a wide range in the disk radial and vertical extents. To obtain reliable rotation curves for the HCN and \ce{C2H} hyperfine lines, we extended standard methodologies to fit multi-component line profiles. We then sampled the likelihood of a thermally stratified model that reproduces all the rotation curves simultaneously, taking into account the molecular emitting layers $z(R)$ and disk thermal structure $T(R,z)$. From this exploration, we obtained dynamical estimates of three fundamental parameters: the stellar mass $M_\star=1.89$ M$_\odot$, the disk mass $M_\text{d}=0.12$ M$_\odot$, and the scale radius $ R_\text{c}=143$ au. We also explore how rotation curves, and consequently the parameter estimates, depend on the adopted emitting layers: the disk mass proves to be the most affected by these systematics, yet the main trends we find do not depend on the adopted parameterization. Finally, we investigated the impact of thermal structure on gas kinematics, and show that the thermal stratification can efficiently explain the measured rotation velocity discrepancies between tracers at different heights. Our results show that such a multi-molecule approach, tracing a large range of emission layers, can provide unique constraints on the ($R,z$) pressure structure of protoplanetary disks.}



\titlerunning{The 2D pressure structure of the HD 163296 protoplanetary disk as probed by multi-molecule kinematics}
\maketitle

\newcommand{\eddyfont}{\ttfamily eddy\rmfamily}
\newcommand{\gofishfont}{\ttfamily GoFish\rmfamily}
\newcommand{\radialprofilefont}{\ttfamily radial{\_}profile\rmfamily}
\newcommand{\SF}[1]{\textcolor{olive}{#1}}
\newcommand{\VP}[1]{\textcolor{orange}{#1}}

\section{Introduction}

In recent years, high spatial and spectral resolution gas observations from the Atacama Large Millimeter/submillimeter Array 
(ALMA) have provided detailed insights into the kinematics and velocity fields of planet-forming disks \citep{pinte2023}. Rotation curves (i.e., the azimuthally averaged radial profiles of the gas rotation velocity) are a direct probe of gas kinematics in protoplanetary disks, and they are crucial to improving our knowledge of several disk properties. 
First, the gas rotation velocity at different heights is determined by the balance between the stellar gravity, the disk self-gravity, and the pressure gradients. As a consequence, accurate rotation curves allow us to dynamically measure the fundamental properties of young star--disk systems, as stellar mass $M_\star$, disk mass $M_\text{d}$, and disk scale radius $ R_\text{c}$. Dynamical masses of protostars were previously estimated by \citet{guilloteau2014, simon2017, yen2018, braun2021}; dynamical measurements of $M_\text{d}$ were instead provided through rotation curves fitting by \citet{veronesi2021} for Elias 2-27, by \citet{lodato} for IM Lup and GM Aur, by \citet{martire} for the whole MAPS sample, and by \citet{andrews2024} for MWC 480. In addition, this methodology was benchmarked and further investigated with models and simulations in recent works by \citet{veronesi2024} and \citet{andrews2024}.
These quantities, especially $M_\text{d}$ and $ R_\text{c}$, are not trivial to measure; rotation curves provide complementary accurate estimates via a methodology that is completely independent of line fluxes or continuum flux densities \citep[see][for a review]{miotello2023}.
Moreover, disks are known to host both radial and vertical gradients in their temperature and density structure \citep[see, e.g.,][for a recent review]{oberg2023}, which jointly determine the pressure structure. With sets of molecular lines originating from different vertical layers, due to a range of abundances, optical depths, and excitation conditions \citep[e.g.,][]{van, aikawa, dartois, semenov,henning, dutrey}, the rotation curves of these same transitions can be leveraged to extract precious information on the pressure structure of disks at different heights.


A first simple approach that we consider in this paper is to verify that the vertical and radial (2D) temperature structure estimated by other means is directly imprinted in the rotation curves of an ensemble of molecular lines spanning a wide range of vertical layers. We focus on temperature rather than density since the former is more directly determined observationally. Historically, observations of both spectral energy distributions and spatially unresolved gas lines in the far-infrared (FIR) and millimeter regimes have been successfully forward-modeled by radiative transfer and sophisticated thermo-chemical codes to infer the thermal structure of the outer regions ($>5-10$ au) of protoplanetary disks \citep[see reviews by][and references therein]{miotello2023,oberg2023}. In recent years, the unique capabilities of ALMA have provided direct stringent constraints on the 2D temperature structure of disks. In particular, with high angular and spectral resolution observations, molecular emitting layers can be reconstructed using geometric arguments from channel maps \citep{pinte2018a}. In the assumption of optically thick and geometrically thin emission, thermalized lines can thus be used to measure the temperature along their emission layer \citep[e.g.,][]{pinte2018a, maps4, paneque}. Using a set of lines originating from different layers, it is thus possible to fit for parameterized 2D ($T(r,z)$) temperature distributions \citep{maps4,law2023,law2024,paneque}. 

Recently, the effect of vertical temperature gradients in rotation curves has been analyzed by \citet{martire}. The authors showed that differences in the rotation velocities of \ce{^12CO} and \ce{^13CO} are not explainable through classical vertically isothermal disk models, but instead must be traced back to vertical stratification in temperature. 

For this work we took a step forward in the analysis of gas kinematics, following a multi-molecule approach. We simultaneously analyzed the rotation velocities of seven molecular tracers emitting from a wide range of disk layers. In this way, our aim is to provide an accurate estimate of the disk mass and scale radius, based on a multi-molecule fit of the curves with a thermally stratified model. From this fit we also investigated the impact of thermal structure on gas kinematics, testing whether vertical stratification can be detected in the kinematic structure of the disk, probed through multiple tracers located at different heights above the midplane. We focused on the very bright disk of HD 163296 and selected exquisite quality multi-molecule data from the Molecules with ALMA at Planet-forming Scales (MAPS) Large Program \citep{maps1}.

This paper is organized as follows. In Sect. \ref{sec:2_observations} we illustrate in detail the selected data and targeted molecular lines. In Sect. \ref{sec:3_methods} we explain the methodology we followed for our analysis. First, we discuss the extracted molecular optical depth profiles that we used as a complementary test to verify the location of the emitting layers. Then, we present the adopted routine for the rotation curves extraction; we describe the new multi-Gaussian methods we implemented to fit the hyperfine structured spectra of \ce{C2H and HCN}, and the corresponding rotation curves. We also highlight a possible observational bias in the rotation curves, arising from beam smearing in the presence of emission gaps. Finally, we describe the model including thermal stratification that we used for the multi-molecule fit of the data. In Sect. \ref{sec:4_results} we show our results from the multi-molecule fit of multiple rotation curves we performed with the thermally stratified model, and we discuss its impact on gas kinematics. From the fit, we derive dynamical estimates of stellar mass $M_\star$, disk mass $M_\text{d}$, and disk scale radius $R_\text{c}$, and we map the 2D pressure structure of the disk. We also discuss the dependence of our results on the assumed emitting surfaces. We finally draw the main conclusions of this work in Sect. \ref{sec:5_discussion_conclusions}.


\section{Observations}
\label{sec:2_observations}
\subsection{The source}
In this work we focus our analysis on HD 163296: thanks to its relative proximity and massive disk, this is one of the most studied Herbig Ae star-disk systems. 
This protoplanetary disk presents multiple features pointing at ongoing planet formation, such as dust rings, azimuthal asymmetries, deviations from
Keplerian velocities, kinks in CO emission, and meridional flows \citep{isella2016, andrews2018, isella2018, teague2019, pinte2018b, izquierdo2022}. Observations in scattered light at optical and infrared wavelengths also show radial substructures \citep{rich}. HD 163296 has been observed in CO and other molecular tracers at millimeter wavelengths in several single-dish and interferometric projects \citep{thi2004, degregorio-mosalvo2013,pegues2020,maps1}. The HD 163296 disk is thought to be hosting two planets in formation, identified by kinematic detections, at 94 au \citep{izquierdo2022} and 260 au \citep{teague2018a, maps18, pinte2018b}. HD~163296 presents extremely bright and radially extended molecular emission, which is pivotal to extract precise and accurate gas kinematics by leveraging the high signal-to-noise ratio (S/N) of specific lines.


\subsection{Targeted emission lines}
To extract information on molecular gas dynamics, we based our work on the publicly available high spatial and spectral resolution data provided by the MAPS ALMA Large Program \citep{maps1}. The survey investigated more than 50 emission lines from different chemical species: in Table \ref{tab:lines} we report the molecules and the corresponding emission lines that we considered for this analysis. An initial visual datacube exploration was performed for each molecule in order to select suitable lines: the candidates were chosen according to their brightness and high signal-to-noise ratio.

\begin{table*}[h!]
    \centering
    \renewcommand\arraystretch{1.2}
    \caption{\centering Molecules and corresponding transitions considered for this work.}
    
    \begin{adjustbox}{max width=\textwidth}
   
    \begin{tabular}{c|c|c|c|c}
    \hline
    \textbf{Molecule} & \textbf{Quantum Numbers} & \textbf{Rest Frequency} & \textbf{ALMA Band} & \textbf{Resolution}\\
    & & \textbf{[GHz]} & & \textbf{[$\prime \prime$]} \\ 
    \hline
        $^{12}$CO &  J=$2-1$ & 230.538000 & 6 & 0.15 \\
        $^{13}$CO & J=$2-1$ & 220.398684 & 6 & 0.15 \\
        C$^{18}$O & J=$2-1$ & 219.560354 & 6 & 0.15 \\
        HCN & J=$3-2$, F=$3-2$ & 265.886434 & 6 & 0.15 \\
        H$_2$CO & J$_\mathrm{K_a,K_b}$=$3_{03}-2_{02}$ & 218.222192 & 6 & 0.3 \\
        HCO$^+$ & J=$1-0$ & 89.188525 & 3 & 0.3 \\
        C$_2$H & N=$3-2$, J=$\frac{5}{2}-\frac{3}{2}$, F=$3-2$ & 262.064986 & 6 & 0.15 \\
    \hline
    \end{tabular}
    \end{adjustbox}
    \tablefoot{For each molecule we report the quantum numbers of the transition, the emission rest frequency, the corresponding ALMA band, and the spatial resolution of the available observations.}
    
    \label{tab:lines}
\end{table*}


\section{Methods}
\label{sec:3_methods}

In this section we give an overview of the general workflow of our analysis and describe in detail the methodology we followed.
Our main aim is to explore the kinematics of the HD 163296 disk through a multi-molecule approach: as different molecules emit at slightly different physical and thermal conditions, considering different emission lines is essential to trace the gas dynamics along the vertical extent of the disk. 
So far, HD 163296 has been kinematically analyzed through its brightest gas emission \citep{pinte2018b, teague2018a, casassus2019, maps4, maps18, izquierdo2023} --\ce{^12CO} and \ce{^13CO} lines; as for abundant tracers with a high signal-to-noise ratio, it is more straightforward to perform a reliable reconstruction of the emitting layer and an exhaustive kinematic analysis.
In this work, we expand the extraction of rotation curves and the following kinematic considerations also to different tracers, to obtain a more complete picture of gas dynamics in the disk.
We show a schematic summary of the workflow:
\begin{itemize}
    \item[$\bullet$] To extract and fit the rotation curves with a thermally stratified model, we assume the 2D thermal structure of the disk from \citet{maps4} and the molecular emitting layers $z(R)$ from \citet{paneque}. We then perform a complementary test to verify the solidity of the extracted emitting layers, given the assumed thermal structure. The validation test consists of the extraction of molecular optical depth profiles $\tau (R)$ from the data: as optical depth is easily related to the height $z(R)$, checking that we obtain reasonable $\tau$ profiles is a good way to validate the solidity of the assumed emitting layers. To extract $\tau$ profiles, we proceed as follows: we consider the temperature structure $T(R,z)$ computed in \citet{maps4} through CO isotopologs and the molecular emitting layers from \citet{paneque}; we extract the molecular optical depth profiles by comparing the surface brightness profile from the data with the expected profile in case of optical thick emission, considering the full Planck law. We assess that our test provides the expected results, in agreement with optical depth estimates obtained with independent methods;
    
    \item[$\bullet$] Once we have verified that the assumed $z(R)$ are reasonable using the optical depth profiles as benchmark, we extract rotation curves from the considered molecules, taking into account the corresponding emitting layers. Considering the large uncertainties on the assumed $z(R)$ from \citet{paneque} (for the faintest lines we consider, $\Delta z/z$ can be as high as $\approx 50\%$), we also check that rotation curves are not significantly influenced by the exact location of the emitting layers, so that the lack of accuracy in $z(R)$ does not severely impact our rotation curves analysis. For the hyperfine lines in our sample (i.e., \ce{HCN} and \ce{C2H}), we extend existing methodologies to more efficiently fit the spectra and obtain improved rotation curves. We also analyze beam smearing effects on curves, verifying that it causes deviations opposite to pressure gradients;
    
    \item[$\bullet$] We fit the obtained rotation curves all together with an extended model, including contributions from the star, the disk self-gravity, the pressure gradient and vertical thermal stratification; from this fit we obtain a dynamical estimate for stellar mass, disk mass, and scale radius. Knowing these parameters and the 2D thermal structure, we show the retrieved 2D pressure structure of the disk. Finally, we compare the predictions from the stratified model and a simple isothermal disk structure, to recover which one can best describe the differences in rotation velocity between the considered molecules. In this way, we show the impact of the pressure structure on the kinematics of molecules different from \ce{^12CO} and \ce{^13CO}, and the improvement induced by the implementation of the vertical stratification, with respect to the isothermal model.
\end{itemize}

\subsection{Molecular optical depth profiles}
\label{subsec:tau}

\citet{maps4} retrieved the emission layers of the three CO isotopologs - \ce{^12CO}, \ce{^13CO}, and \ce{C^18O} - within the radial extent where they are optically thick. In the optically thick regime, the gas temperature along the emitting layers can be easily measured from the peak surface brightness of the emission lines \citep{weaver2018}; thus, from the lines profiles in each pixel it is possible to extract the gas kinetic temperature along the CO surfaces, assuming LTE conditions. They fitted the obtained temperature values with the 2D parametric expression for the kinetic temperature of the gas by \citet{dullemond2020}, which depends on the radius $R$ and the height above the midplane $z$:
\begin{equation}
    T^4(R,z)=T^4_\mathrm{mid}(R)+\frac{1}{2} \left[ 1+\tanh{\left(\frac{z-\alpha z_q(R)}{z_q(R)}\right)}\right] T_\mathrm{atm}^4(R).
    \label{eq:T_law}
\end{equation}
Here $z_q(R)=z_0 (R/100\mathrm{\ au})^\beta$, $T_\mathrm{atm}=T_\mathrm{atm,0}(R/100\mathrm{\ au})^{q_\mathrm{atm}}$, and $T_\mathrm{mid}=T_\mathrm{mid,0}(R/100\mathrm{\ au})^{q_\mathrm{mid}}$. They fitted seven parameters using data from the three CO optically thick isotopologs: $T_\mathrm{atm,0}$, $q_\mathrm{atm}$, $T_\mathrm{mid,0}$, $q_\mathrm{mid}$, $\alpha$, $z_0$, and $\beta$. We report the best-fit parameters obtained by \citet{maps4} in Table \ref{table:law_parameters}. In this work, we use the kinetic temperature by \citet{maps4} and the observed brightness temperature from the data to derive radial profiles for the molecular optical depth.

\begin{table}[h!] 
    \centering
    \renewcommand\arraystretch{1.2} 
    \caption{Best-fit values for the seven parameters found in Eq. \ref{eq:T_law}, as fitted by \citet{maps4} through CO optically thick isotopologs.}
    \label{table:law_parameters}
    \begin{adjustbox}{max width=\columnwidth}
    \begin{tabular}{c|c}
    \hline
        \textit{$T_\mathrm{atm,0}$} [K] & 63 \\ 
        \textit{$q_\mathrm{atm}$} & -0.61 \\ 
        \textit{$T_\mathrm{mid,0}$} [K] & 24 \\
        \textit{$q_\mathrm{mid}$} & -0.18 \\
        \textit{$\alpha$} & 3.01 \\
        \textit{$z_0$} [au] & 9 \\
        \textit{$\beta$} & 0.42 \\
    \hline
    \end{tabular}
    \end{adjustbox}
\end{table}

We also consider \ce{H2CO} ($3_{03}-2_{02}$), \ce{HCO+} ($J=1-0$), HCN ($J=3-2, F=3-2$), and \ce{C2H} ($N=3-2, J=5/2-3/2, F=3-2$). We adopt the molecular emitting layers by \citet{paneque} into Eq. \ref{eq:T_law}, to find the kinetic temperature of the gas at the specific layers. With the obtained temperature profile, we calculate the expected surface brightness profile $I_\mathrm{Planck}$ at the emitting layer, under the assumption of optically thick emission, through the Planck law:
\begin{equation}
    I_\mathrm{Planck}(\nu, T) = \frac{2h\nu^3}{c^2} \frac{1}{e^{\frac{h\nu}{k_\mathrm{B}T(R,z)}}-1}.
\end{equation}
Here $\nu$ is the rest frequency of the considered line and $T(R,z)$ is the temperature on the emitting surface computed from Eq. \ref{eq:T_law}. 
Then, we extract the effective surface brightness $I_\mathrm{data}$ from the data, using the \gofishfont \ \citep{Teague_GoFish} function \radialprofilefont, where we set a $30$° wedge along the disk major axis.
At this point, we know that the two obtained profiles for the surface brightness are linked by the following relation:
\begin{equation}
    I_\mathrm{data}(R) = I_\mathrm{Planck}(R) \left[ 1-e^{-\tau(R)}\right].
    \label{eq:tau}
\end{equation}
By inverting Eq. \ref{eq:tau}, we can compute the optical depth radial profile $\tau (R)$ as
\begin{equation}
    \tau(R) = - \ln{\left(1-\frac{I_\mathrm{data}(R)}{I_\mathrm{Planck}(R)}\right)}.
\end{equation}

\begin{figure*}
   \centering
   \includegraphics[width=\textwidth]{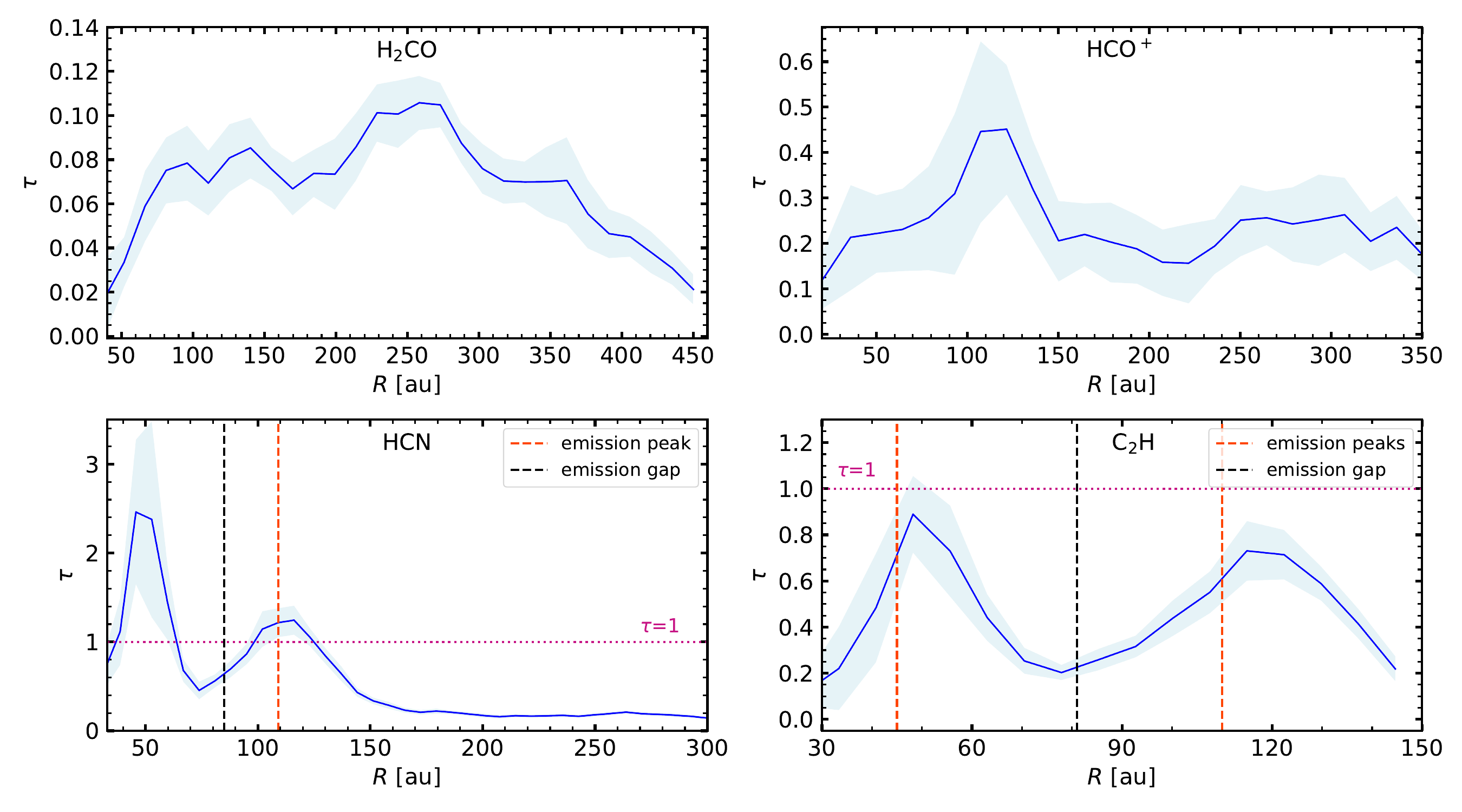}
      \caption{
      Optical depth radial profiles for \ce{H2CO}, \ce{HCO^+}, HCN, and \ce{C2H} found in this work, taking into account the molecular emitting layers. Emission peaks and emission gaps, when present, are highlighted in red and black dashed lines, respectively. We also report the $\tau =1$ line for easier visualization.}
         \label{fig:tau_profiles}
\end{figure*}

In Fig. \ref{fig:tau_profiles} we show the optical depth profiles of \ce{H2CO}, \ce{HCO^+}, HCN, and \ce{C2H} we obtained following this method. We calculated uncertainties on the $\tau$ profiles (shaded profiles in Fig. \ref{fig:tau_profiles}) propagating the errors on the extracted surface brightness. Uncertainties are generally larger for \ce{H2CO} and \ce{HCO^+}, the two lines with lower resolution data ($0.3''$).

First of all, we observe that \ce{H2CO} and \ce{HCO^+} lines result in an optically thin emission along the whole radial extent of the disk, reaching maximum peaks at $\tau \sim 0.11$ and $\tau \sim 0.45$ respectively. \ce{C2H} emission becomes marginally thick, with a value for $\tau$ approaching to 1, at the radii corresponding to intensity peaks. 
HCN emission is mostly optically thick within $\sim 130$ au, with higher values in the inner $\sim 60$ au and a local maximum around the radius of the intensity peak ($R=109$ au); it then evolves into optically thin emission, with $\tau$ decreasing by an order of magnitude. As expected, \ce{C2H} and HCN result in a marginally thick emission at the rings locations, as previously observed in other disks \citep{maps1,oberg2023}. 
We also find that in the innermost tens au we measure a general decrease of the optical depth for all the molecules. This decrease is an observational artifact and is due to the beam dilution effect \citep{weaver2018}: at smaller scales, the beam size may be larger than the spatial dimension of the considered emitting region of the disk. As a consequence, the measured surface brightness is underestimated by a beam dilution factor, proportional to the ratio of the emitting area to the beam size (lower than 1 at small radii).

The obtained estimates for $\tau$ are compatible with existent optical depth values measured for the same molecules in this \citep{maps11} and other disks \citep{bergner2019,facchini2021}, and are in line with the expected optical depth values that can be retrieved from column density profiles in \citet{maps6} and from the brightness temperature profiles computed by \citet{paneque} for HD 163296.
Therefore, the consistent results of this preliminary test guarantee that, given the thermal structure for the disk by \citet{maps4}, the emitting layers by \citet{paneque} are suitable for the rotation curves extraction. Considered the large errors on the assumed molecular emitting layers, we also verified that the rotation curves obtained with \eddyfont \ are not significantly affected by the emitting layers choice, varying them within the given uncertainties. Thus the assumed layers are suitable to extract reliable curves, without the need to rely on deeply accurate measurements of $z(R)$, whose extraction procedure shows high systematic uncertainties and is thus intrinsically difficult.

Finally, we note that this is an efficient and straightforward method to easily extract optical depth and column density profiles of the considered line, relying only on thermal structure $T(R,z)$ and emitting height $z(R)$ of one single transition. This is an independent method that does not require complicated chemical studies about excitation temperatures and represents a valid alternative to approaches applied to hyperfine lines \citep{maps6} for systems where a precise and accurate 2D temperature structure is known.

\subsection{Rotation curves extraction}
\label{subsec:curves_extraction}
Once we have verified with the optical depth complementary test described in the previous section that the assumed emitting layers are consistent, we proceed to the rotation curves extraction.
In order to extract the velocity radial profiles of the HD 163296 disk, we used the \eddyfont \ Python package \citep{teague_eddy}, following the method presented in \citet{teague2018a} and \citet{teague2018b}. First of all, we have to know the geometry of the disk (i.e., inclination $i$, position angle $PA$, possible disk center offsets in declination and right ascension) and the emitting layers of each considered molecule, to deproject the data from projected location in the sky to disk-frame coordinates. We assumed $i=46.7$°, $PA=313.3$° from \citet{maps1} (the $PA$ is shifted by $180$° according to its definition in \eddyfont) and center offsets equal to zero. For the molecular emission layers we assumed an exponentially tapered power law
\begin{equation}
    z(R) = Z_0 \left(\frac{R}{100\text{au}}\right)^\phi \exp\left[-\left(\frac{R}{R_{\mathrm{taper}}}\right)^{\psi}\right],
\end{equation}
with best-fit values by \citet{paneque}, listed in Table \ref{table:emitting_layers}, for $Z_0, \phi, r_{\mathrm{taper}}$, and $\psi$. In addition, we also performed our analysis adopting nonparametric emitting layers for all the molecules, and we show the results we obtained in Appendix \ref{app:nonpar_layers}. In Sect. \ref{subsec:layers_dependency} we discuss these results, showing how the choice of using parametric emitting surfaces for the considered lines has no substantial effect on the qualitative results and trends we obtain from our analysis.

\begin{table}[h!] 
    \centering
    \renewcommand\arraystretch{1.2} 
    \caption{\centering Best-fit parameters for the molecular emitting layers for HD 163296 fitted by \citet{paneque}.}
    \label{table:emitting_layers}
    \begin{adjustbox}{max width=\columnwidth}
    \begin{tabular}{c|c|c|c|c}
    \hline
     & \textbf{$Z_0$ [au]} & \textbf{$\phi$} & \textbf{$r_{\mathrm{taper}}$ [au]} & \textbf{$\psi$} \\
    \hline
        \textbf{$^{12}$CO} & 25.08&1.35 &426.44 &4.21 \\ 
        \textbf{$^{13}$CO} &11.88 &1.4 &343.88  &5.36 \\ 
        \textbf{C$^{18}$O} &6.02 &1.47 &/ &/ \\
        \textbf{H$_2$CO} &19.2 &1.0 &462.85  &0.07  \\
        \textbf{HCO$^+$} &6.45 &0.59 &/  &/  \\
        \textbf{HCN} &8.45 &2.14 &342.11  &3.75  \\
        \textbf{C$_2$H} &38.13 &5.41 &95.38  &3.78  \\
    \hline
    \end{tabular}
    \end{adjustbox}
    \tablefoot{For C$^{18}$O and HCO$^+$ a simple power law is assumed as parameterization for the emitting layer, so only $Z_0$ and $\phi$ are reported.}
    
\end{table}
We note that taking into account the molecular emitting layers in the data deprojection does not influence the extracted curves in a dramatic way, for molecules emitting close to the midplane. Especially we are interested in testing the molecules with lower S/N emission, for which it is more difficult to extract reliable rotation curves (i.e., \ce{H2CO, HCO^+, C2H}). To test this, we computed the rotation curves of these molecules assuming flat emission (i.e., originating at the disk midplane) and we obtained that the differences between the two scenarios are compatible with zero within: $0.6\sigma$ for \ce{HCO^+ and H2CO}, $4.5\sigma$ for \ce{C2H} (and $3\sigma$ for \ce{C^18O}). For thoroughness, we did the same test for all the molecules: as expected, for the other lines the discrepancy is above $6\sigma$, as these molecules are well known to emit from very elevated layers. For every line, we show the difference between the extracted rotation curves in case of flat and elevated emission in Fig. \ref{fig:flat_emission_diff}.
We also note that the curves would be even more compatible if we had more realistic estimates for the uncertainties on the rotation curves, as the errors on the curves are underestimated, as we explain in Sect. \ref{subsec:multimol_fit}.
This test guarantees a reliable extraction of rotation curves for lower S/N lines emitting close to the midplane, even if the available parameterizations for their emitting heights have large uncertainties, without affecting the quality of the results. In Sect. \ref{subsec:layers_dependency} we show the results on the stellar and disk parameters we obtain from the multi-molecule fit of the rotation curves in case of flat emission for all the considered molecules.

To extract the rotation curves, we used the \textit{SHO (Simple Harmonic Oscillator)} method in \eddyfont, that assumes axisymmetric disks. Following this approximation, emission line spectra that arise from the same radial distance from the disk center are expected to exhibit the same profile, as they trace the same physical and chemical conditions. The line centroids are expected to be shifted by $v_\theta \cos(\theta)$, an offset with respect to the systemic velocity due to the azimuthal location of the emission in the disk ($\theta$ represents the polar angle, measured from the redshifted major axis).
With this method, the best-fit rotation velocity at a specific radius is determined as the one that best aligns the deprojected spectra from the same annulus. Mathematically, we fit the set of line centroids of the same annulus with a cosine function.
In order to align the spectra, in our analysis we fit each spectrum with a Gaussian curve to identify the intensity peak and the corresponding velocity, and then we proceed to the fit of the line centroids with the harmonic function.
For each tracer we repeat the extraction of the rotation curve $20$ times, considering different independent samples of spectra inside each annulus for different iterations and we compute the uncertainties on the velocity values as their weighted standard deviation.
A similar approach is followed in \citet{maps18} to extract the \ce{^12CO} rotation curve for the HD 163296 disk with \gofishfont: in that case, the best-fit velocity values are selected as those that best align the spectra, as quantified by fitting a Gaussian processes mean-model to the averaged spectra.

If we want to consider also the radial contribution to the line-of-sight velocity $v_{\mathrm{los}}$, in \eddyfont \ we can fit the azimuthal and radial velocities together, and fitting every set of line centroids with a double harmonic oscillator. 
We note that for molecules other than the CO, the MAPS quality of data is not sufficient to provide reliable kinematical estimates of the radial and vertical velocity components.  
Nevertheless, in our analysis we are interested in the azimuthal velocity profiles. We decided not to focus on $v_R$ and $v_z$, considering them negligible with respect to the difference in azimuthal velocities between data and an axisymmetric model, as shown by \citet{teague2019} and \citet{izquierdo2023}.

\subsubsection{Limitations of the method}

As explained in Sect. \ref{subsec:curves_extraction}, the extraction of rotation curves is based on the identification and fit of line centroids in each pixel of the data cube. Besides requiring a high signal-to-noise ratio, this methodology relies on a few assumptions. First, we assumed that the disks are azimuthally symmetric and that the spectra along each annulus vary only by a phase $\cos \theta$, but maintain the same profile. This hypothesis might be inadequate for highly perturbed line emission morphologies.
Second, \eddyfont \ does not distinguish the upper and lower surfaces of disks, and assumes that the front side dominates over the back side.
However, an important contamination from the back side of the disks can complicate the spectra alignment and so the goodness of the fit. In some cases (such as GM Aur in the MAPS sample, see \citealt{lodato}) the signals arising from the two surfaces are comparable and thus the procedure results in a quite inefficient alignment of the spectra. In these cases, the lower surface can shift the line centroids in the observed intensity spectra, inducing a wrong derivation of the best-fit velocities. We note that the level of contamination from the back side depends on the considered disk and molecule: for example, line emission coming from lower layers is generally less contaminated by the back side with respect to the emission from the upper layers, since its emission surface is closer to the disk midplane \citep{lodato}.

To quantify the effects of these limitations on our analysis, we quantitatively compared the rotation curves we extracted with \eddyfont \ with the ones obtained by \citet{martire} with the \texttt{discminer} tool \citep{izquierdo2021} for \ce{^12CO} and \ce{^13CO}, considering also the back side of the disk. We measure very low differences, which are $<4\%$ for \ce{^12CO} and $<5\%$ for \ce{^13CO}, confirming that the limitations of this method are not our primary source of systematic effects.

\subsubsection{Multi-Gaussian methods for hyperfine transitions kinematics: C$_2$H and HCN rotation curves}
\label{subsec:hyperfine}

\begin{figure*}[h!tbp]
    \centering
    \includegraphics[width=.45\textwidth]{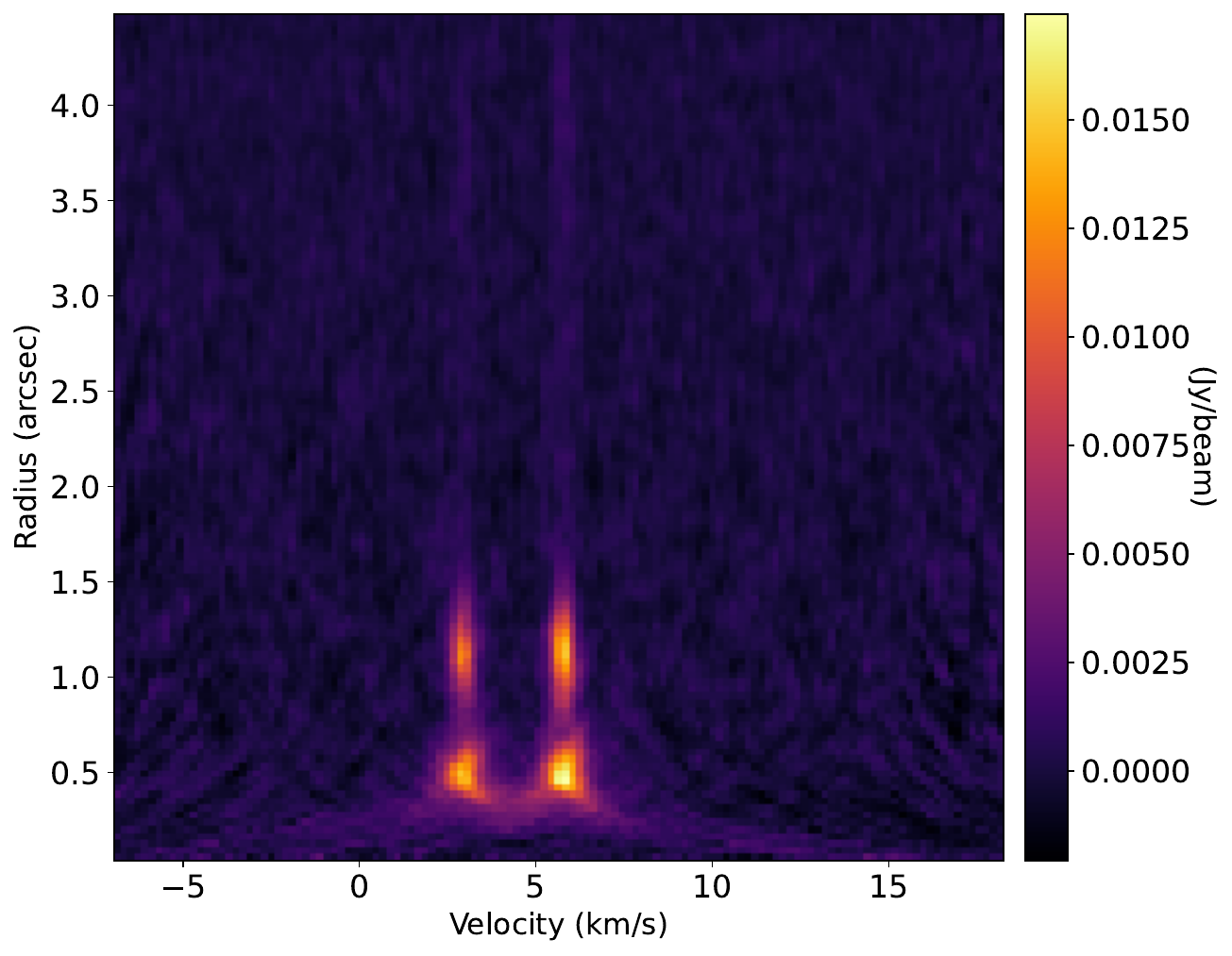}
    \includegraphics[width=.45\textwidth]{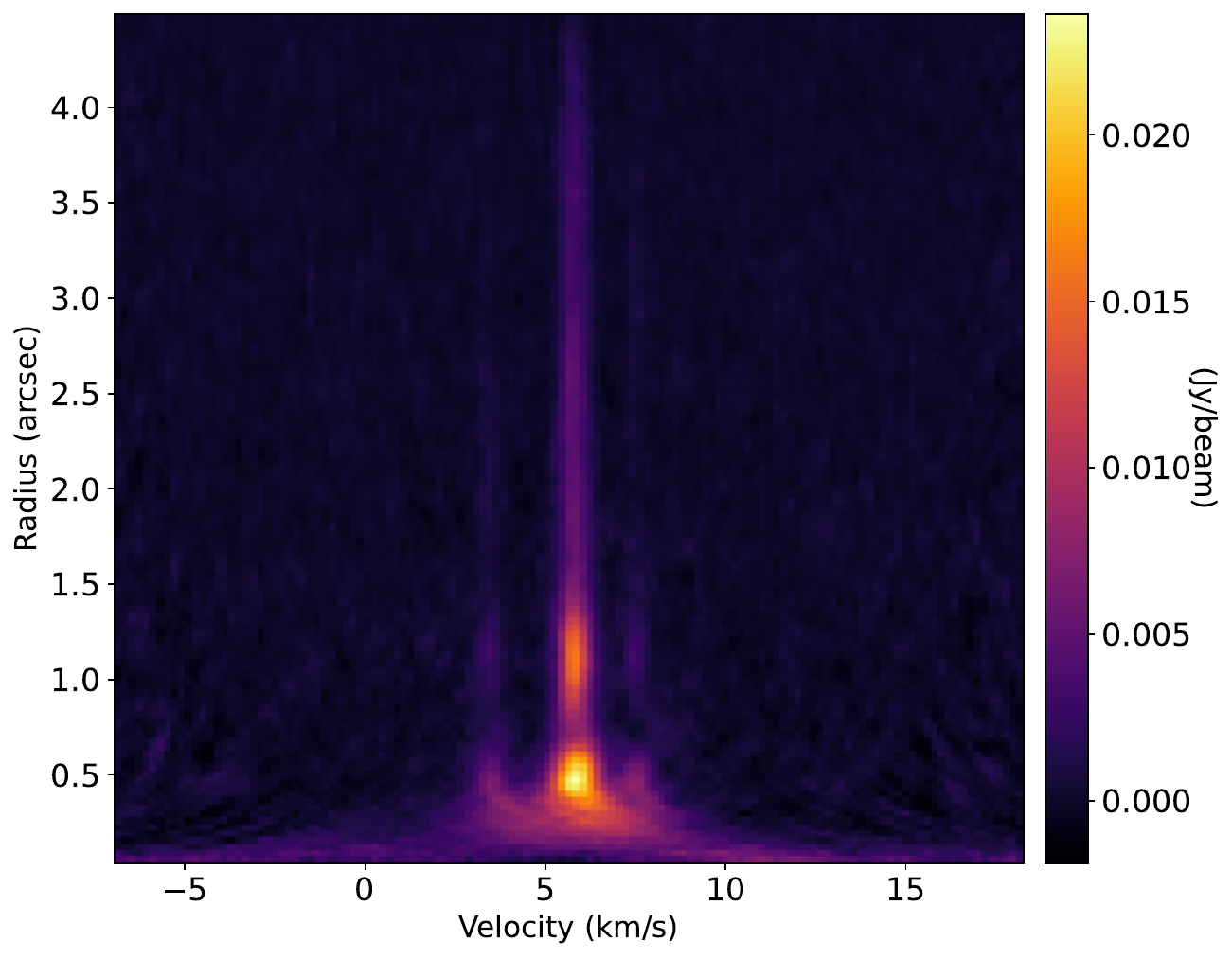}
    \caption{
    Teardrop plots for \ce{C2H} (left) and HCN (right). Each row of the plot represents an averaged spectrum at separate annuli after shifting and stacking. The color scale indicates the intensity of the emission line. In both cases, hyperfine transitions (two for \ce{C2H} and three for HCN) are clearly visible.}
    \label{fig:teardrops}
\end{figure*}

In this work, we developed a new method to extract rotation curves from molecules that have hyperfine transitions.
In fact, in our analysis molecules such as C$_2$H and HCN do exhibit a hyperfine structure (see Table \ref{tab:lines} for details about the transitions). Hyperfine structure is typically characterized by energy shifts that are orders of magnitude smaller than fine-structure ones. Therefore, for these molecules, it is possible to detect two (or more) emission lines in the same frequency channel (i.e., in the same channel map). 
Typically, a spectrum from molecules with hyperfine lines shows two or more different intensity peaks (one for each transition) instead of one, as clearly shown by teardrop plots in Fig. \ref{fig:teardrops}, where the color scale represents the averaged intensity of the considered emission line along separate annuli. The intensity of each visible transition depends on its Einstein coefficient: for \ce{C2H}, in each spectral channel we see two similar transitions, roughly equal in intensity. This reflects in double-peaked spectra with two maxima at the same intensity. For HCN, instead, we can clearly detect a central dominant transition, which is the strongest one, alongside with two fainter lines nearly of the same intensity. Therefore, the standard extraction methods (that fit a quadratic or Gaussian function to the spectrum) are not adequate to fit these multi-peaked spectra, and they fail to return reliable velocity profiles. 

For this reason, we implemented in \eddyfont \ a new fitting procedure for hyperfine lines using multiple Gaussian components (two or three, depending on the number of hyperfine transitions), where the width of the multiple Gaussians is assumed to be the same, as the hyperfine components are expected to emit under the same conditions of temperature, density, and turbulence.

For \ce{C2H}, we fit each spectrum with a double-Gaussian function
\begin{equation}
    I = I_0\ \exp \left[{\left( \frac{v-v_0}{\sigma}\right) ^2}\right] + I_0 \ \xi \ \exp \left[ {\left( \frac{v-(v_0-\Delta v)}{\sigma}\right) ^2}\right],
\end{equation}
where $\xi$ is the ratio of the two intensity peaks and $\Delta v$ is the velocity difference between the two transitions.
We fit four free parameters: the center of the Gaussian $v_0$, which represents the line centroid, the width of the Gaussians $\sigma$, the intensity corresponding to the first peak $I_0$, and the peak intensity ratio $\xi$.

For \ce{HCN}, we fit each spectrum with a triple-Gaussian function
\begin{equation}
\begin{split}
    I = &\ I_0 \ \xi_\mathrm{blue}\ \exp \left[{\left( \frac{v-(v_0-\Delta v_\mathrm{blue})}{\sigma}\right) ^2}\right] + I_0\ \exp \left[{\left( \frac{v-v_0}{\sigma}\right) ^2}\right] \\ 
    &+ I_0 \ \xi_\mathrm{red} \ \exp \left[ {\left( \frac{v-(v_0+\Delta v_\mathrm{red})}{\sigma}\right) ^2}\right],
\end{split}
\end{equation}
where $\xi_\mathrm{blue}$ ($\xi_\mathrm{red}$) is the ratio of the blueshifted (redshifted) intensity peak to the central dominant peak and $\Delta v_\mathrm{blue}$ ($\Delta v_\mathrm{red}$) is the velocity difference between the central and the blueshifted (redshifted) peak.
In this case we fit five free parameters: the center of the middle Gaussian $v_0$, which represents the line centroid, the width of the Gaussians $\sigma$, the intensity corresponding to the central peak $I_0$, and the peak intensity ratios $\xi_\mathrm{blue}$ and $\xi_\mathrm{red}$.

The velocity shifts $\Delta v$ (between the two \ce{C2H} hyperfine lines), $\Delta v_\mathrm{blue}$, and $\Delta v_\mathrm{red}$ (between the weaker hyperfine lines and the central one of HCN) are fixed parameters, given the known frequency shifts $\Delta \nu = 2.5$ MHz (between the two \ce{C2H} transitions), $\Delta \nu_\mathrm{blue} = 2.09$ MHz, and $\Delta \nu_\mathrm{red} = 1.54$ MHz (between the HCN transitions) as taken from \textit{the Cologne Database for Molecular Spectroscopy (CDMS)} (\citealt{sastry1981,muller2000} for \ce{C2H} and \citealt{ahrens2002} for HCN). We set priors for some parameters: 
\begin{itemize}
    \item[$\bullet$] Intensities must be positive: $I_0,\ \xi,\ \xi_\mathrm{blue},\ \xi_\mathrm{red} >0$;
    \item[$\bullet$] $\xi \in [0.2,\ 1.5]$ for \ce{C2H} and $\xi_\mathrm{blue},\ \xi_\mathrm{red} \in [0.1,\ 1.]$ for HCN: we choose a reasonably large interval centered on the expected value for the ratio between the intensity peaks, considering the Einstein coefficients and upper state degeneracies of the considered transitions ($\xi \sim A_{ij}^1g_1/A_{ij}^2g_2 \approx 0.65$ for \ce{C2H} and $\xi \sim A_{ij}^1g_1/A_{ij}^2g_2 \approx 0.13$ for HCN);
    \item[$\bullet$] $\sigma \in [0,\ \Delta v /2]$ for \ce{C2H} and $\sigma \in [0,\ \Delta v_\mathrm{blue} /2]$ for HCN: the width of the Gaussians is not larger than half of the distance in velocity between the two considered transitions, as the expected thermal broadening of the line is below $\sim 1$ km/s \citep{izquierdo2023} and thus much smaller than $\Delta v$.
    
\end{itemize}
For \ce{C2H}, we set reasonable starting positions for the parameter space exploration, to ease the convergence of the fit. In particular, we computed the Keplerian velocity at each radius $R$ and azimuthal angle $\theta$ as
\begin{equation}
    v_{\mathrm{rot, K}} = \Biggl(\frac{GM_{\filledstar}}{R}\Biggr)^{1/2} \sin i \ \cos \theta + v_\mathrm{sys},
\end{equation}
where $M_{\filledstar}$ is the stellar mass computed fitting the \ce{^12CO} curve with a Keplerian model, and used it as starting position for the line centroid $v_0$ fit.

\begin{figure}[h!tbp]
    \centering
    \includegraphics[width=\columnwidth]{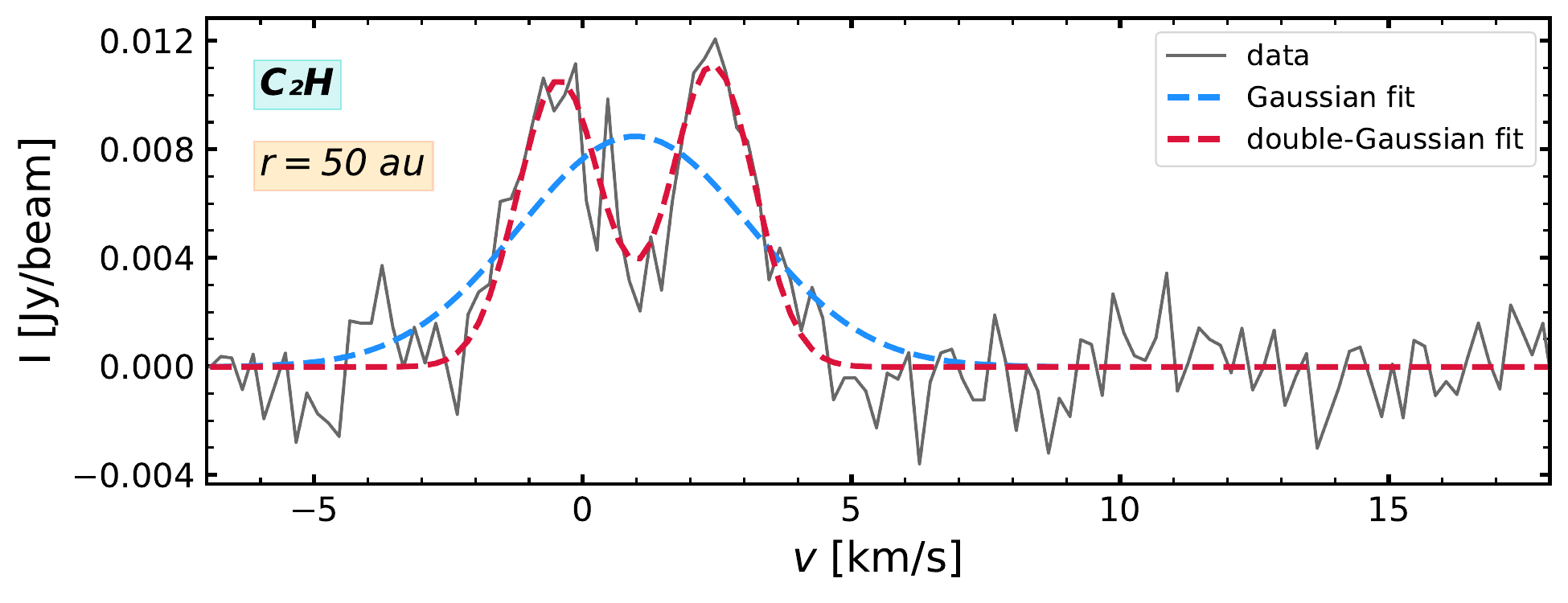}
    \includegraphics[width=\columnwidth]{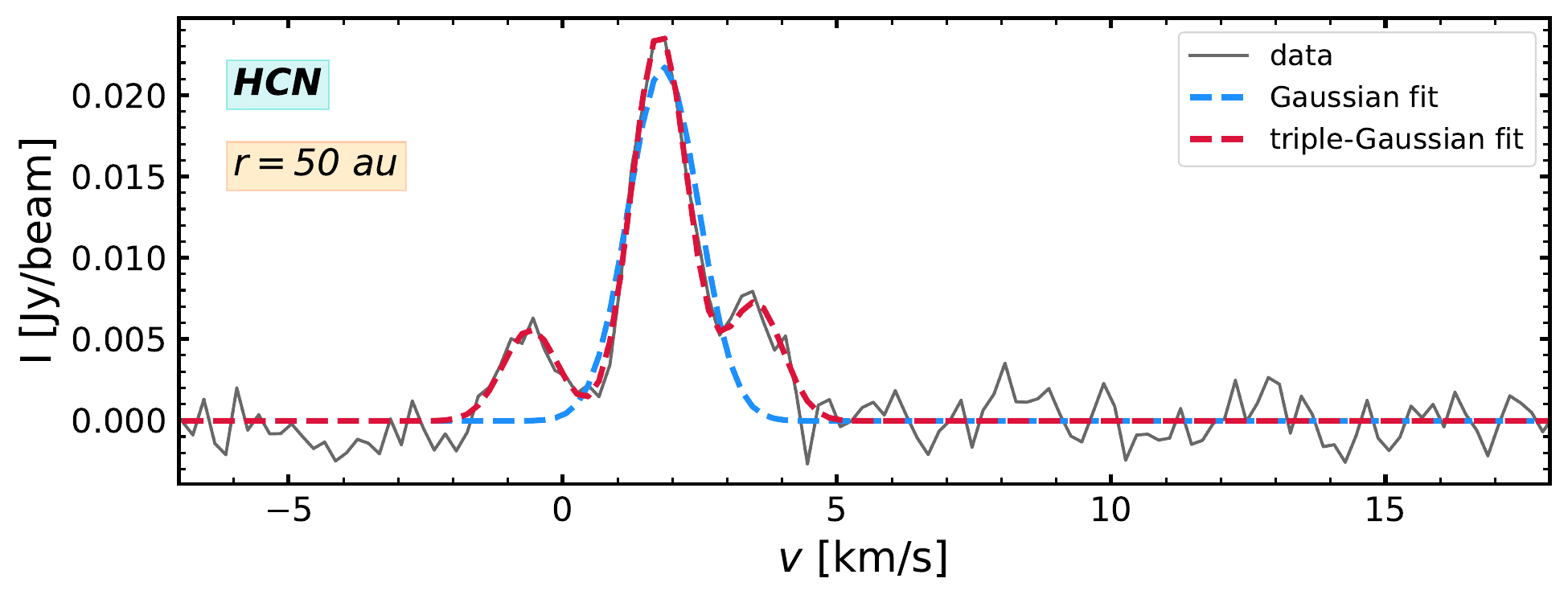}
    \caption{
    Improvement in the fitting of the spectra using multi-Gaussian methods.\\
    \textit{Top panel}: Comparison between the single-Gaussian (light blue) and the double-Gaussian (pink) methods used to fit the intensity peaks of C$_2$H at $50$ au. \textit{Bottom panel}: Same as top panel, but for the three hyperfine components of the HCN transition.}
    \label{fig:hyper_spectra}
\end{figure}

\begin{figure}
   \centering
   \includegraphics[width=\columnwidth]{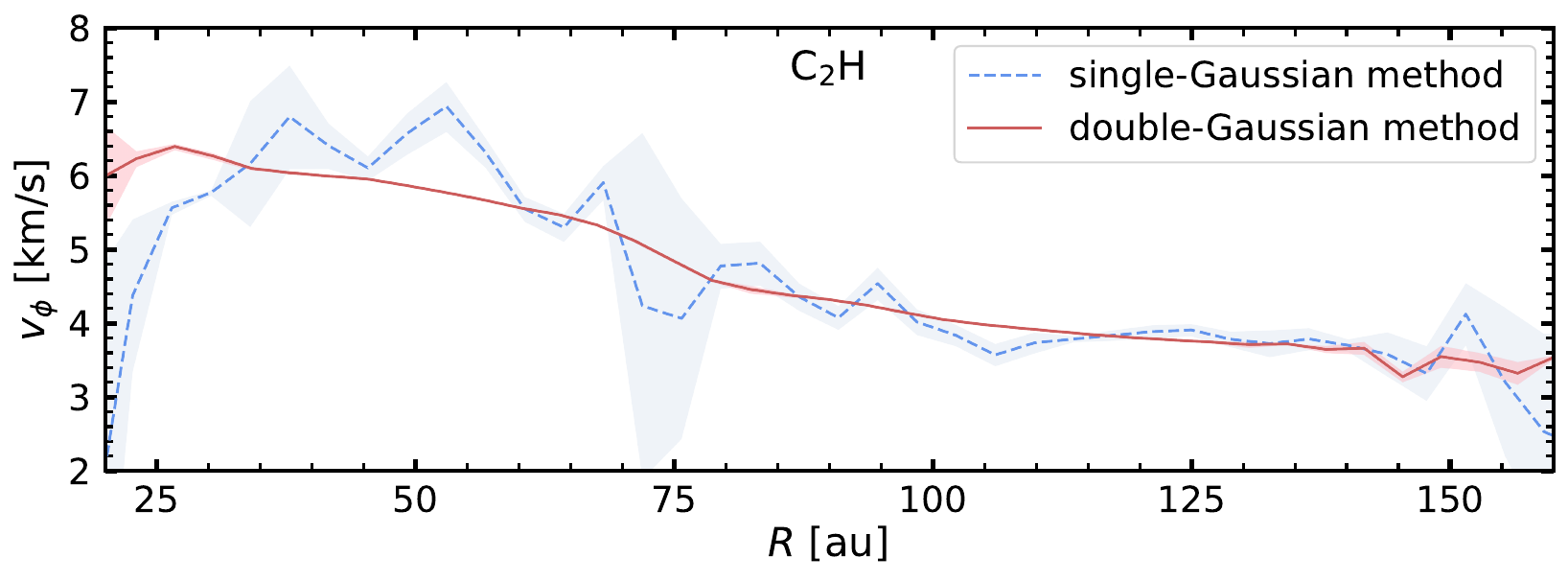}
   \includegraphics[width=\columnwidth]{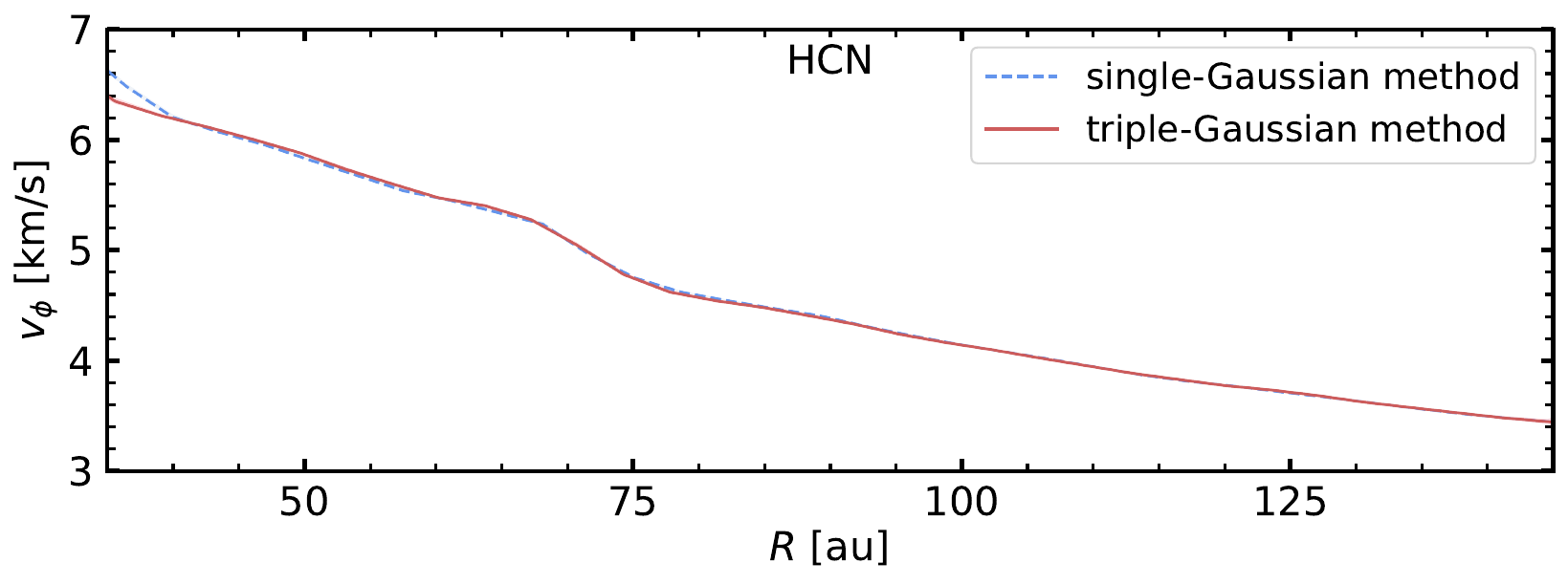}
      \caption{
      Comparison between the rotation curves for \ce{C2H} (top) and HCN (bottom) using the double- or triple-Gaussian methods (pink) and the single-Gaussian method already implemented in \eddyfont \ (dashed light blue) for the fitting of the spectra.}
         \label{fig:hyperfine_curves}
\end{figure}

In Fig. \ref{fig:hyper_spectra} we show the fit of the hyperfine spectra of both \ce{C2H} and HCN with, respectively, the double- and triple-Gaussian methods, compared to the single-Gaussian one originally implemented in \eddyfont.
These methods lead to a remarkable improvement in the extraction of the rotation curves of hyperfine structured lines. In Fig. \ref{fig:hyperfine_curves}, for both molecules we show the comparison between the rotation curve extracted with the multi-Gaussian method and the curve previously obtained with the single-Gaussian method already implemented in \eddyfont.

These methods prove very efficient in fitting the multi-peaked spectra in both cases. For \ce{C2H}, the remarkable upgrade in the peak fitting procedure leads to the most significant improvement in the azimuthal velocity profile: fitting the spectra with a double-Gaussian, we obtain a smooth curve with small errors, a more reliable result than the one obtained using a single-Gaussian. In that case, since the two peaks in the \ce{C2H} spectra are mostly equal in intensity, a single-Gaussian model leads to a fragmented curve affected by high variance.
For HCN, by fitting three Gaussian components to each spectrum we maximize the exploitation of the available data, taking into account also the two minor intensity peaks. Nevertheless, the two curves do not differ substantially and the observed differences in the selected radial range are small: as the central peak is dominant, also the standard single-Gaussian method, even if less accurate than the triple-Gaussian, could still fit the central prominent peak providing the correct value of the line centroids, neglecting the hyperfine splitting visible in the two minor peaks.


\subsubsection{Beam smearing effect}
\label{subsec:beam_smearing}

Among the considered molecules, \ce{C2H} and HCN show strong intensity gradients in their emission. Since the extraction of velocity profiles comes from molecular line emission, the presence of gradients in the emission intensity of these lines can affect the corresponding rotation curves. In Fig. \ref{fig:anticorr}, we show these observational effects for \ce{C2H}; we show the same effect for HCN in Appendix \ref{app:HCN}.

We find a strong anticorrelation between the two following measurable quantities as a function of radius (evaluated at the same radii), in agreement with the results from \citet{keppler2019}:
\begin{itemize}
    \item[$\bullet$] $\dd\ {(\int{Idv})}/\dd{R}$: the intensity gradient of \ce{C2H} molecular emission line, computed from the radial intensity profile\footnote{Radial intensity profiles are available in the MAPS database at \url{https://alma-maps.info/data.html}.};
    \item[$\bullet$] $v_\mathrm{C_2H}-v_{\mathrm{^{12}CO}}$: the difference between the rotation velocities of \ce{C2H} and $^{12}$CO (shown separately in panel (b) of Fig. \ref{fig:anticorr}). In this case, the $^{12}$CO curve was chosen as reference because it exhibits a typical smooth Keplerian profile, and it does not show any strong gradient in the intensity radial profile. 
\end{itemize}

These two quantities display an opposite modulation, as shown in panel (c) of Fig. \ref{fig:anticorr}. In particular, the difference in rotation velocity $v_{\ce{C2H}}-v_{\ce{^12CO}}$ is positive at the inner edge of an emission gap, while it becomes negative at its outer edge. This effect is shown in the comparison between the curves reported in panel (b) of Fig. \ref{fig:anticorr}.
\begin{figure}[h!tbp]
    \centering
    \includegraphics[width=\columnwidth]{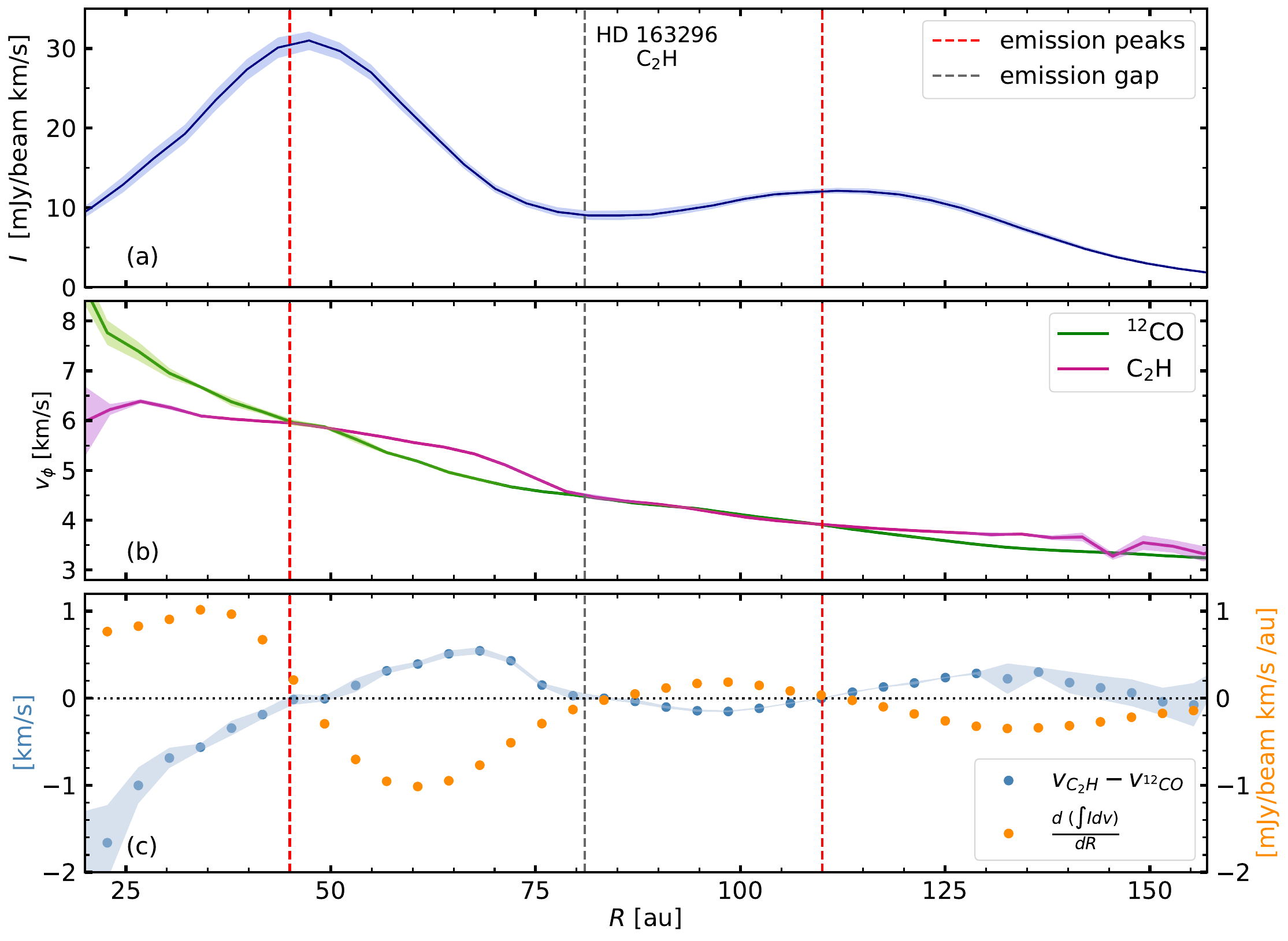}

    \caption{
    Beam smearing effect for \ce{C2H}.\\
  \textit{Panel (a)}: Azimuthally averaged integrated intensity profile of the C$_2$H line. Emission peaks are marked with dashed red lines and the emission gap is highlighted with a dashed grey line. \textit{Panel (b)}: Rotation curve of C$_2$H (purple line) compared to that of $^{12}$CO (green line). The rotation velocity of C$_2$H is super-Keplerian just before the gap and sub-Keplerian after. It shows the opposite behavior in correspondence of the emission peaks. \textit{Panel (c)}: Anticorrelation between the C$_2$H integrated intensity gradient and its difference in rotation velocity compared to $^{12}$CO. The two quantities show an opposite modulation, and they cross at the radii corresponding to the emission gap or peaks.}
    \label{fig:anticorr}
\end{figure} 
This particular modulation can be explained by a beam smearing effect. When a source is observed at a given angular resolution, every extracted spectrum is the intensity weighted average of all the spectra that fall inside the beam area. For this reason, a strong intensity gradient across the beam would bias the recovered spectrum and, consequently, the extracted velocity. In our results we are biased to see velocities corresponding to higher intensity values inside the beam, that do not correspond to the beam center. 
For instance, in correspondence of the inner edge of the emission gap we are affected by a strong negative intensity gradient. So in this region, the extracted spectrum weighs more the velocities corresponding to the higher intensities, which are the ones from the inner radii, and we measure an artificially higher rotation velocity. On the contrary, at the outer edge of the emission gap the strong intensity gradient is positive, and velocities corresponding to the outer radii are weighted more and we measure an artificially lower rotation.

Beam smearing can also explain the flattening at small radii we see in the \ce{C2H} rotation curve in panel (b) of Fig. \ref{fig:anticorr}, since this molecule shows a central cavity in emission, followed by an intensity peak. In the internal cavity, the signal inside the beam is dominated by the higher intensities found at larger radii: as a consequence, we measure lower velocities in correspondence of the cavity, leading to a flattening of the curve.

\begin{figure}[h!tbp]
    \centering
    \includegraphics[width=\columnwidth]{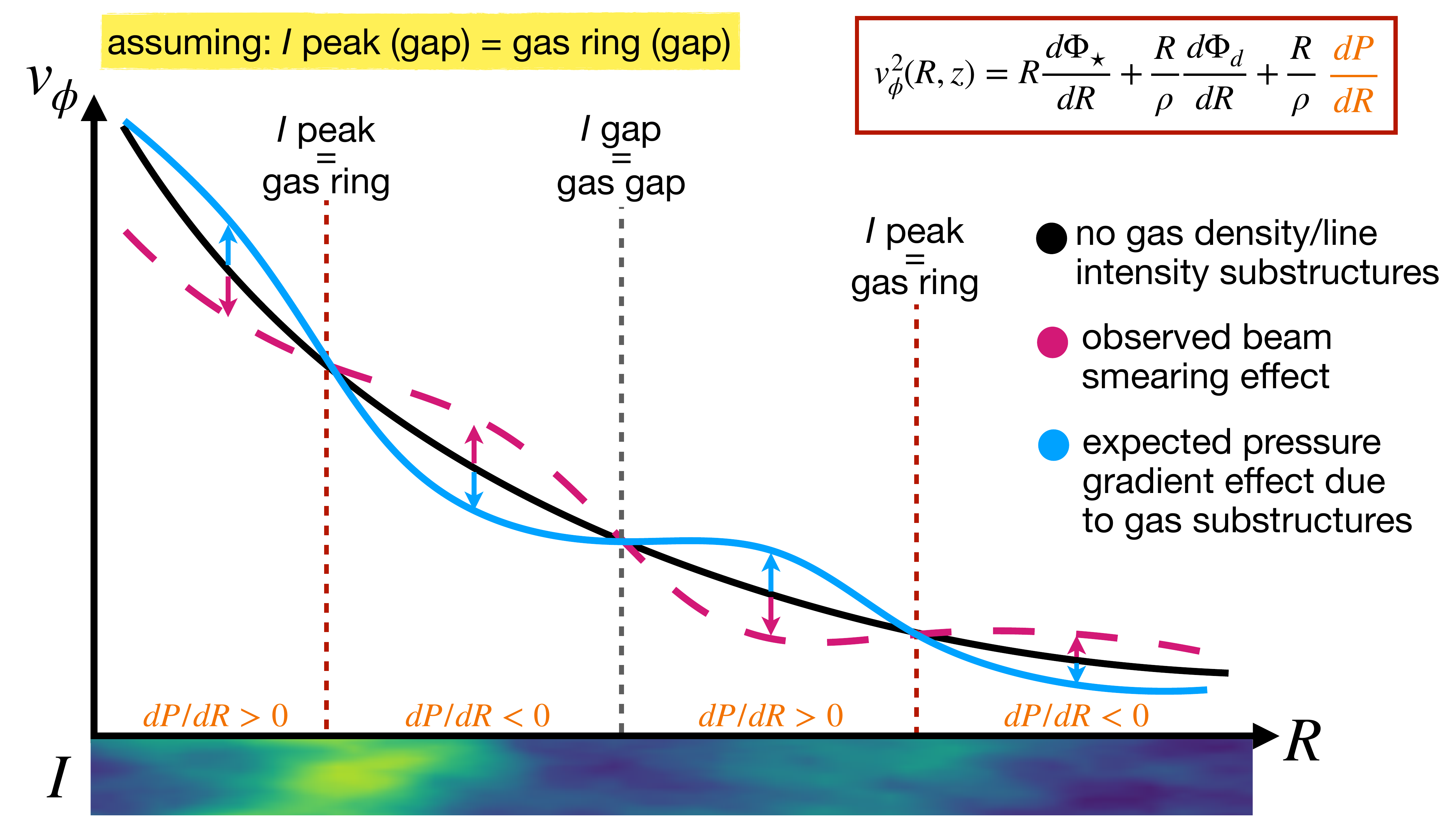}

    \caption{Sketch of the different effects that beam smearing due to line intensity gradients (pink dashed line) and additional pressure gradient effects due to the presence of gas density substructures (light blue solid line) have on a standard rotation curve (black solid line). We assume that intensity gradients trace gas density gradients; a peak or gap in line intensity corresponds to a peak or gap in gas density. 
    Line intensity and gas density peaks are marked with dashed red lines, and line intensity and gas density gaps are highlighted with a dashed grey line. At the bottom we show a colored map of the radial intensity. The solid black line represents the rotation curve of a molecule showing no gas density or line intensity substructures. The typical curve distortion due to beam smearing effect is shown by the pink dashed line, and is in agreement with our observations. The expected distortion of the curve due to the presence of gas density substructures is shown by the light blue solid line. The rings and gaps in the gas density induce additional pressure gradients that influence the rotation curve, according to the $v^2_\phi(R,z)$ expression shown in the upper right corner. These pressure gradients are positive before rings and beyond gaps locations, while they are negative before gaps and beyond rings locations. Thus, in this case, the deviations in rotational velocity are expected to be in the opposite direction of the beam smearing-induced variations.}
    \label{fig:sketch_beam_smearing}
\end{figure}

This observational effect has important consequences: we have observed that beam smearing triggers apparent velocity deviations in rotation curves, that are linked to strong intensity gradients of the emission line. Nevertheless, if any substructure (such as rings or gaps) is present in the gas surface density, then also the pressure profile undergoes alterations induced by the change in density, causing similar velocity deviations. Therefore, it is important to distinguish rotation curve modulations that are due to this observational bias from effective deviations linked to pressure gradients, that could be signs of the presence of gas density substructures. However, we note that beam smearing works in an opposite way to pressure gradients, as we show in Fig. \ref{fig:sketch_beam_smearing}: assuming that intensity peaks trace pressure peaks (and analogously with gaps), pressure gradients are expected to induce velocity deviations that are of opposite sign with respect to the ones deriving from intensity gradients that we observe. 
This means that the apparent difference in velocity induced by beam smearing in correspondence of an intensity gap could be incorrectly interpreted as a pressure maximum, not as a pressure gap as expected if intensity and pressure peaks correspond.
Conversely, if we observe kinematic evidence of a pressure gap in correspondence of an intensity gap, we can immediately conclude that it cannot have been artificially generated by beam smearing, since this effect would have induced the opposite behavior (and analogously for a pressure peak). This different impact that beam smearing and pressure gradients have on rotation curves may help distinguish these effects; nevertheless, this may be not straightforward, since both of them can be present and influence data in different ways.

\subsection{Vertically stratified model}
We extract and analyze multiple molecular lines emitting from different layers to investigate how disk kinematics is affected by height above the midplane across the vertical extent of the disk. Recent high resolution gas observations have shown differences in the rotation velocity of the CO isotopologs, that could not be correctly explained with simple vertically isothermal disk models. In \citet{martire}, the authors show that a thermal stratified model can better explain the velocity discrepancies between the \ce{^12CO} and \ce{^13CO} rotation curves, and that vertical stratification should be considered when building a solid model to adequately describe them. Therefore, we choose to simultaneously fit our seven rotation curves including the contribution from vertical thermal stratification, in order to provide a complete picture of the several contributions that can affect the disk kinematics. From this fit we derive dynamical estimates for the disk and stellar masses, and the disk scale radius; we also compare our results with a simple vertically isothermal model, showing that it is not adequate to describe the complexity of the data.

We consider a disk surface density as described by the self-similar solution of \citet{lyndenbell}:
\begin{equation}
\label{eq:surf_density}
    \Sigma = \frac{(2-\gamma)M_\text{d}}{2\pi R_\text{c}^2} \Bigg(\frac{R}{R_ \text{c}}\Bigg)^{-\gamma}\exp\Bigg[-\Bigg(\frac{R}{R_\text{c}}\Bigg)^{2-\gamma}\Bigg].
\end{equation}
Here $M_\text{d}$  and $ R_\text{c}$ are the disk mass and the scale radius, $R$ is the cylindrical radius, and $\gamma$ is the index that describes the steepness of the surface density (we adopt $\gamma=1$). On the midplane we have that the temperature can be described by a power law as $T_\mathrm{mid}(R) \propto R^{\ q_\mathrm{mid}}$ and consequently the sound speed is $c_{\text{s, mid}} \propto \sqrt{T_\text{mid}} \propto R^{\ q_\mathrm{mid}/2}$.
To account for the thermal vertical stratification of the disk, following the method and parameterization described in \citet{martire}, we can define two functions $f$ and $g$ that determine the vertical dependence of the disk temperature, sound speed, and density:
\begin{gather}
    \label{eq:temp} T(R,z)=T_\text{mid}(R)f(R,z),\\ \label{eq:cs} c_\text{s}^2(R,z)=c_{\text{s, mid}}^2(R)f(R,z), \\ \rho(R,z)=\rho_\text{mid}(R)g(R,z).
\end{gather} 
As a consequence, considering a barotropic fluid, the 2D pressure in the disk is given by
\begin{equation}
    \label{eq:pressure} P(R,z)=P_\text{mid}(R)f(R,z)g(R,z) = c_{\text{s, mid}}^2(R) \rho_\text{mid}(R)fg(R,z).
\end{equation}
We note that $g(R,z)$ is a function that only depends on $f(R,z)$ as a consequence of the hydrostatic equilibrium (see \citealt{martire}).

We adopt the model for our rotation curves fit from \citet{martire}, given by
\begin{equation}\label{eq:rotationcurve_strat}
\begin{split}
    v_\theta^2(R,z) = &v_\text{k}^2 \left\{ \left[1+\left(\frac{z}{R}\right)^2\right]^{-3/2} - \left[ \gamma^\prime + (2-\gamma)\left(\frac{R}{R_\mathrm{c}}\right)^{2-\gamma} \right. \right. \\
    &\left. \left. - \frac{\text{d}\log(fg)}{\text{d}\log R}\right]\left(\frac{H}{R}\right)_\text{mid}^2 f(R,z) \right\} \\
    &+ G \int^\infty_0 \Bigg[K(k) - \frac{1}{4}\Bigg(\frac{k^2}{1-k^2}\Bigg)\times \\
    &\Bigg(\frac{r}{R}-\frac{R}{r}+\frac{z^2}{Rr}\Bigg) E(k)\Bigg]\sqrt{\frac{r}{R}} k\Sigma(r) \,dr,
\end{split}
\end{equation}
where $v_\text{k}=\sqrt{GM_\star/R}$ is the Keplerian velocity, $\gamma^\prime = \gamma + (3+q)/2 $, $r=\sqrt{R^2+z^2}$ is the spherical radius, $K(k)$ and $E(k)$ are complete elliptical integrals \citep{abramowitz}, and $k^2=4Rr/[(R+r)^2+z^2]$.
In this expression we can distinguish the different contributions of the stellar gravity, the pressure gradient, the thermal stratification (the term depending on $f$ and $g$), and the self-gravity \citep{bertin}.
For further details on this model, see \citet{martire}. The adopted expression for the isothermal model is explained in \citet{lodato}, Eqs. 12-14.

To perform the simultaneous fit of the rotation curves, we use the code DySc\footnote{The code is publicly available at \url{https://github.com/crislong/DySc}.}: given the data, the thermal structure of the disk, and the geometry of the emitting layers, the code performs the fit of the data to the model by maximizing the likelihood of the model, with the posterior distribution explored by Monte Carlo Markov Chains, as implemented in emcee\rmfamily\footnote{For details, we refer to \citet{emcee}.}. As output, we obtain the probability distributions of the fitted parameters, their best-fit values, and corresponding uncertainties.
We implemented the possibility to fit multiple ($>2$) rotation curves in the code. We fit our seven rotation curves obtained from data with a single model for the rotation velocity described in Eq. \ref{eq:rotationcurve_strat}. From this multi-molecule fit, we can extract accurate best-fit values for some fundamental parameters: stellar mass $M_\star$, disk mass $M_\text{d}$, and scale radius $ R_\text{c}$.

\section{Results and discussion}
\label{sec:4_results}

\subsection{Multi-molecule fit with thermal stratification: dynamical estimates and pressure structure}
\label{subsec:multimol_fit}

We extended the code DySc to perform a multi-molecule fit with seven different molecules and we fitted the obtained molecular rotation curves simultaneously with one single model, including contributions from the stellar and the disk gravity, the pressure gradient and thermal stratification, as described by Eq. \ref{eq:rotationcurve_strat}. We ran the MCMC exploration using 200 walkers, 1000 burn-in steps and 2000 steps to determine the final posterior distribution of the fitted parameters. 
To evaluate the contributions of vertical structure, we took into account the molecular emitting layers found by \citet{paneque}.
For the thermal structure to use in our fit, we considered the parameterization for the temperature presented by \citet{dullemond2020} and reported in Eq. \ref{eq:T_law}, with the parameters obtained by \citet{maps4} reported in Table \ref{table:law_parameters}. In this work we did not account for the uncertainties in the temperature structure of the disk. \citet{andrews2024} analyzed the systematic uncertainties driven by the disk thermal structure, with related uncertainties that are of the same order of surfaces mis-specifications.

\begin{figure*}
   
   \includegraphics[scale=0.3]{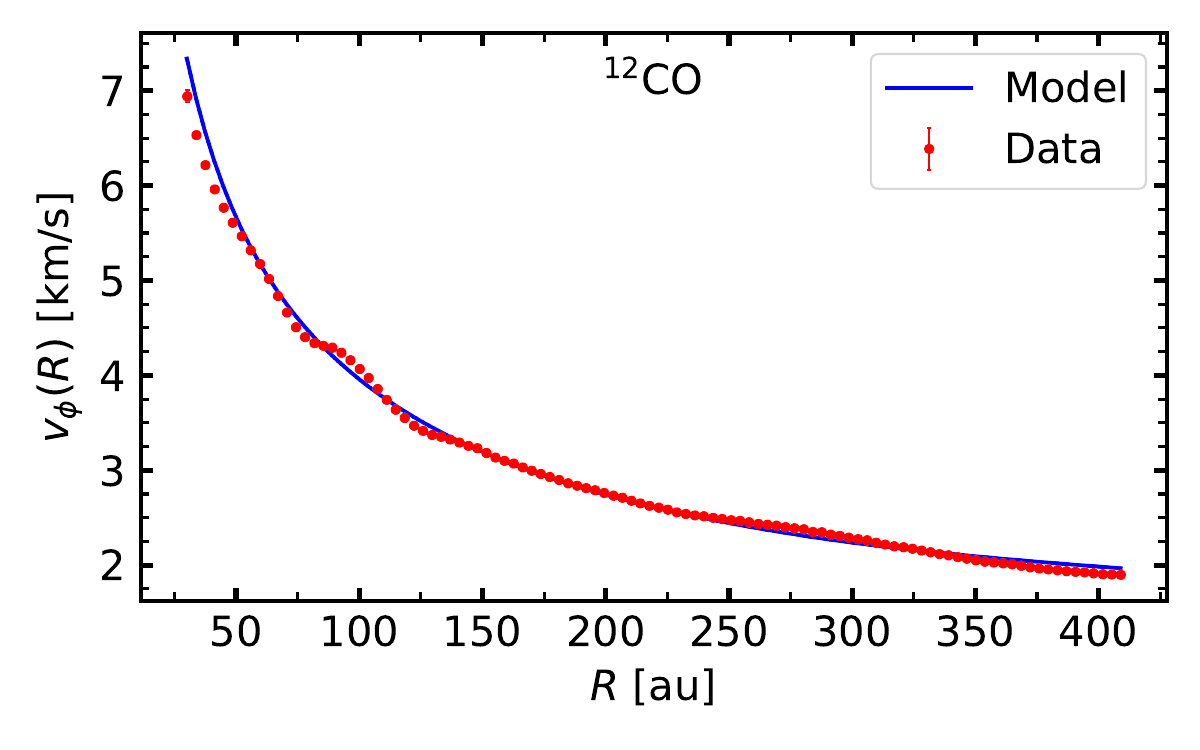}
   \includegraphics[scale=0.3]{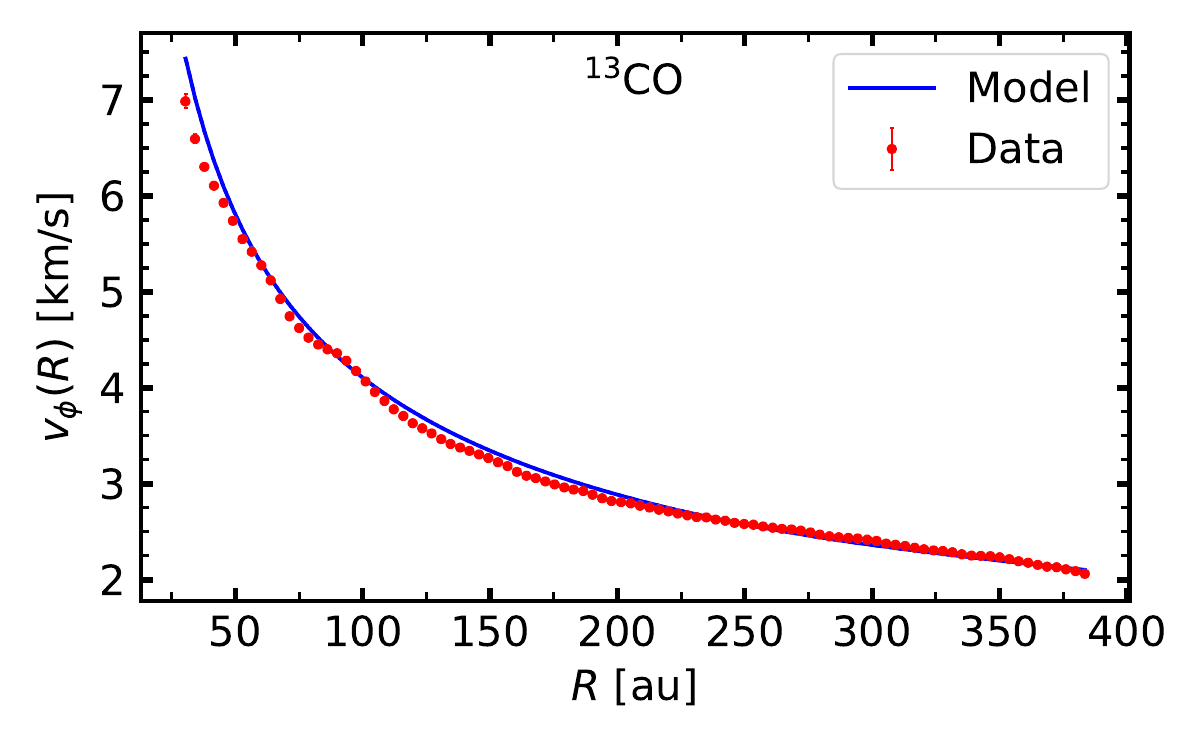}
   \includegraphics[scale=0.3]{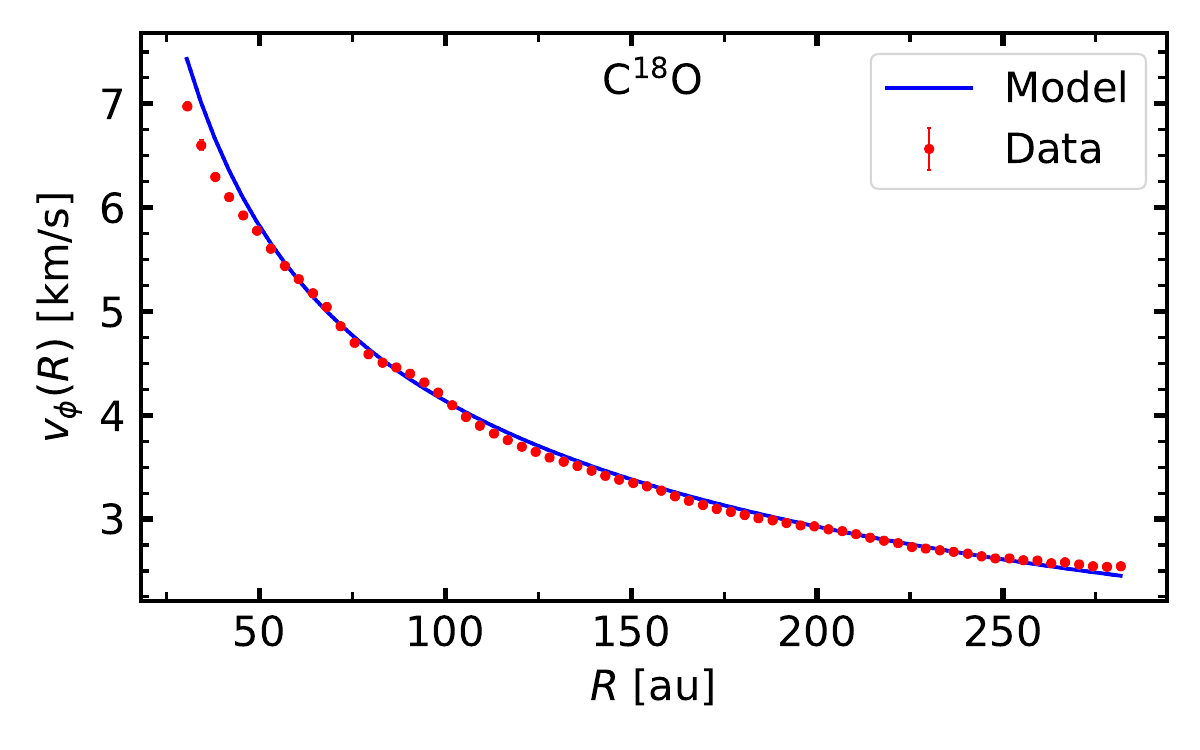}
   \includegraphics[scale=0.3]{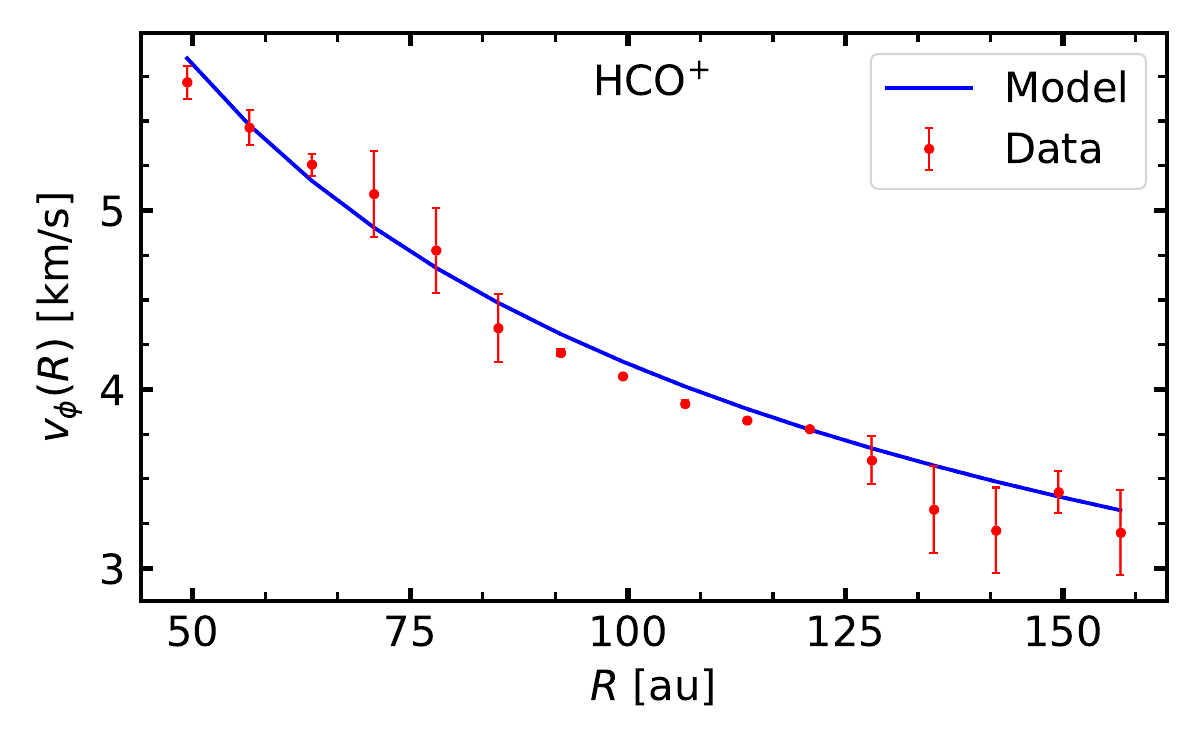}
   \includegraphics[scale=0.3]{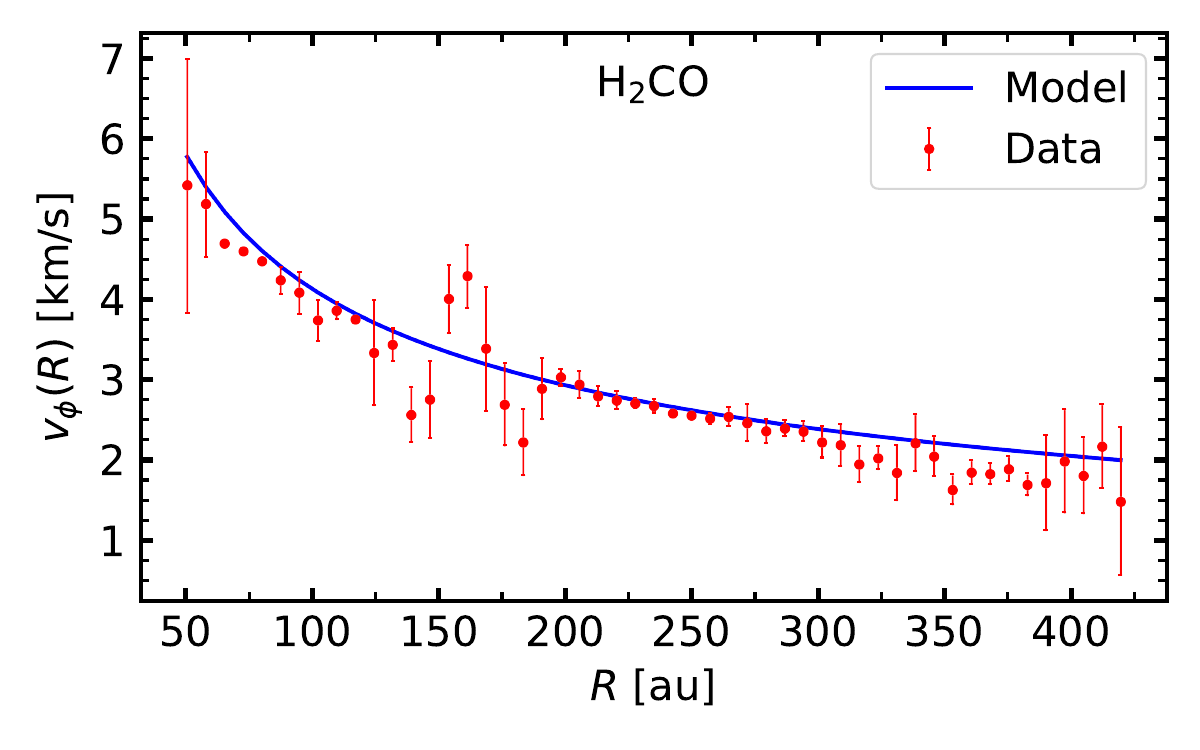}
   \includegraphics[scale=0.3]{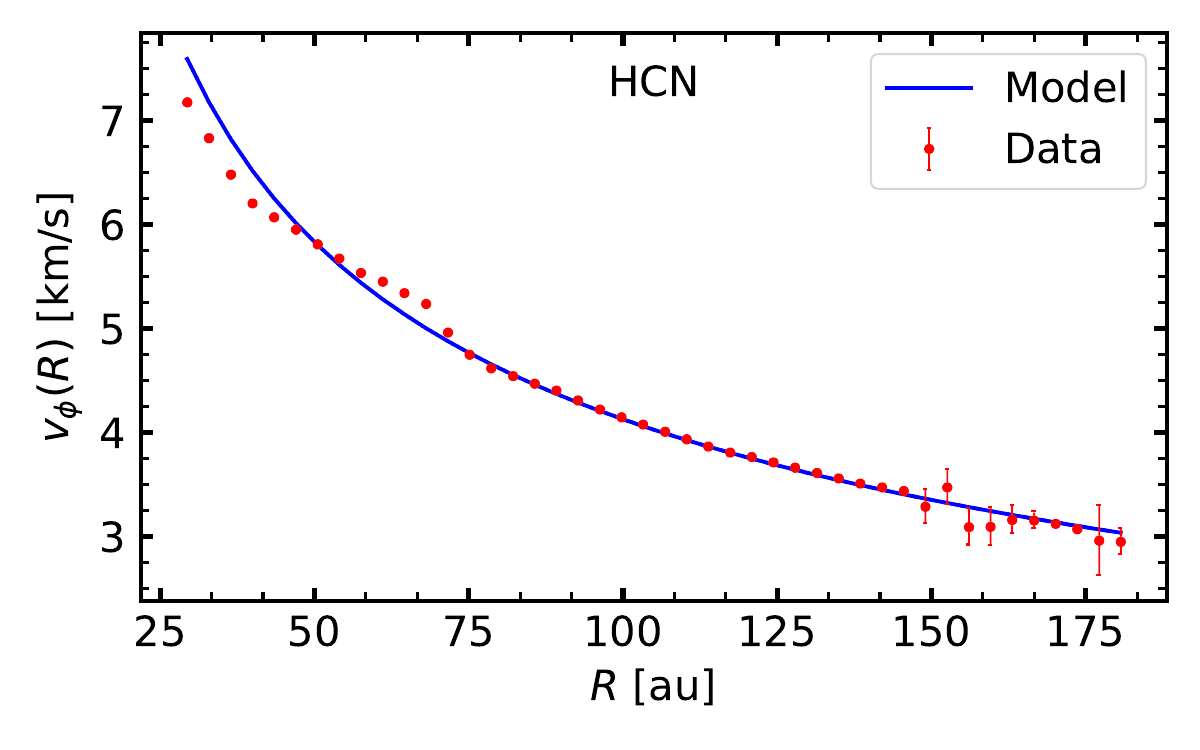}
   \begin{center}
        \includegraphics[scale=0.3]{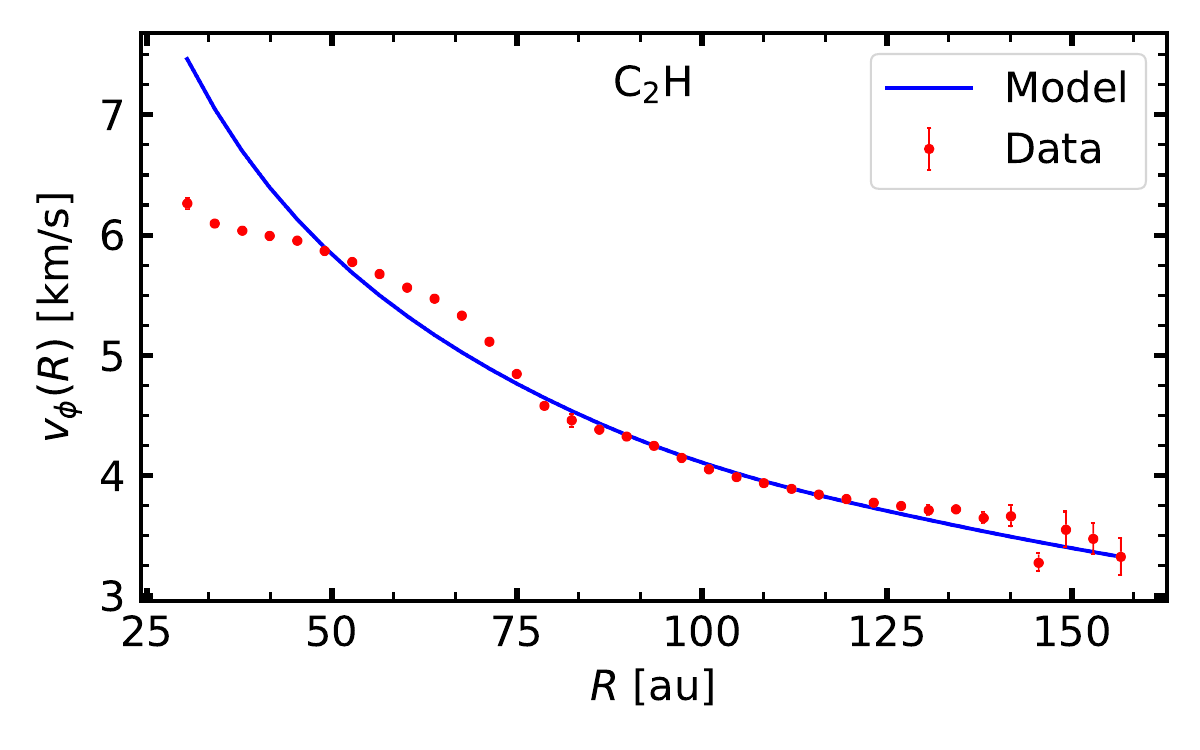}
   \end{center}
      \caption{
      Multi-molecule fit of the seven considered rotation curves with the thermally stratified model. For each tracer, the red dots with error bars represent the data, while the blue lines are the best-fit models of each velocity profile.}
         \label{fig:multimolecular_fit}
\end{figure*}

We report the multi-molecule fit in Fig. \ref{fig:multimolecular_fit}: we show the fit of each molecular rotation curve with the recovered best-fit model. It is clear that the model well reproduces the seven curves, with the exception of a short range of radii in \ce{H2CO} around $R \sim 150$ au: in a short interval around this radius, which represents an emission gap for \ce{H2CO}, the reconstruction of the rotation curve with \eddyfont \ is not accurate, since the signal-to-noise ratio is too low to allow an efficient extraction of the rotation velocity (see Appendix \ref{app:H2CO} for further details). As for HCN and \ce{C2H}, the differences with respect to the model are due to the apparent deviations in the rotation curves caused by beam smearing, as explained in Sect. \ref{subsec:beam_smearing}. 
It is important to note that the inclusion of several molecules different from the CO isotopologs, with their corresponding emitting heights, does not hinder the successful outcome of the fit, but rather it contributes to its accuracy.

The uncertainties on the rotation curves are very small: this is an underestimate as we are taking into account only the statistical errors on the values, but no systematic errors. To provide a more exhaustive estimate of the uncertainties, it would be necessary to implement in \eddyfont \ the systematic component of the errors on the derived rotation curves associated with the uncertainty on the geometrical parameters of the disk (inclination, position angle, and centering), the thermal structure, and the molecular emitting layers. \citet{andrews2024} discuss an efficient way forward in the determination of systematic uncertainties in kinematic measurements.

From the multi-molecule fit, we are able to find the best-fit values for the stellar mass $M_\star$, the disk mass $M_\text{d}$, and scale radius $ R_\text{c}$. We note the potential of detailed kinematical analysis in extracting dynamical estimates of these crucial quantities, whose estimate may be very difficult with other methods. 
For each parameter, we compute a 95\% credibility interval (that is the Bayesian counterpart of the confidence level) as the range of values that contains 95\% of the probability density in the posterior distribution and we consider the extremes of these intervals as estimates of the best-fit parameters uncertainties.
We summarize the best-fit values we obtained from the multi-molecule fit, including the thermal stratification contribution, in Table \ref{table:dyn_estimates}; for completeness, we report the corresponding corner plots in Appendix \ref{app:cornerplot}. We compare these estimates with the values extracted from a multi-molecule fit, this time assuming an isothermal temperature profile $T(R) \propto R^{\ q}$ with $H/R=0.084$ and $q=-0.84$ \citep{maps5}. We also compare our results with the ones obtained by \citet{martire}: the authors retrieve the rotation curves using \texttt{discminer} \footnote{See \citet{izquierdo2021} for details.}, and use only \ce{^12CO} and \ce{^13CO} for the fit, including thermal stratification contribution.

\begin{table}[h!]\caption{\centering Comparison of best-fit values.}\label{table:dyn_estimates}
    \centering
    \renewcommand\arraystretch{1.2} 
    \begin{adjustbox}{max width=\columnwidth}
    \begin{tabular}{c|c|c|c}
    
    \hline
    & stratified - 2 mol & stratified - 7 mol & isothermal - 7 mol \\
    & \citep{martire} & (this work) & (this work) \\
    \hline
        \textbf{M$_\filledstar$} [M$_\odot$] & $1.948 \pm 0.002$ & $1.8905 \pm 0.0002$ & $1.8905 \pm 0.0002$ \\ 
       
        \hline
        \textbf{M$_\mathrm{d}$} [M$_\odot$] & $0.134 \pm 0.001$ & $0.1161 \pm 0.0003$ & $0.1565 \pm 0.0003$ \\
        
        \hline
         \textbf{R$_\mathrm{c}$} [au] & $91 \pm 1$ & $143.1 \pm 0.1$ & $48.1 \pm 0.1$ \\
        
    \hline
    \end{tabular}
    \end{adjustbox}
    \tablefoot{Comparison between best-fit values for $M_\star$, $M_\text{d}$, and $ R_\text{c}$ with the corresponding uncertainties obtained by \citet{martire} and from this work, with the thermally stratified and the isothermal models.}
\end{table}

    
       
        
        

We note that in our work the estimate of the stellar mass does not change between the isothermal and stratified fits, while the disk mass appears slightly lower in the second scenario. What changes the most is the disk scale radius, as expected: in the isothermal case, we obtain $R_\mathrm{c}=48.1$ au, a very low value for the scale radius and a clear sign that we are probably underestimating it. On the contrary, with the thermally stratified model we provide a more accurate characterization of the thermal structure of the disk and this leads to $R_\mathrm{c}=143.1$ au. This value is more consistent with the best-fit estimates of $R_\mathrm{c}$ obtained with complementary methods by \citet{degregorio-mosalvo2013} and \citet{maps5}. 
When comparing our results for the thermally stratified scenario with the ones from \citet{martire}, we find that they obtain slightly higher values for both the stellar and the disk masses, and a lower estimate of the disk scale radius. We point out that the fitting procedure in the two works is the same; the main differences are represented by the rotation curves to fit (the authors use \texttt{discminer} by \citealt{izquierdo2021} to extract the velocity profiles, while we use \eddyfont) and the number of considered emission lines (they use \ce{^12CO} and \ce{^13CO} while we use seven different molecules).
We also observe that we obtain uncertainties on the best-fit parameters that are an order of magnitude smaller than the ones obtained by \citet{martire}. This is understandable considering that \eddyfont \ and \texttt{discminer} are based on different methods to provide rotation curves, so we do not expect to obtain the same order of magnitude for the uncertainties a priori. Error bars on the determined velocity values affect the simultaneous fit performed on the curves with DySc: as we are underestimating the uncertainties on the rotation curves with \eddyfont, consequently we obtain low estimates of the errors on the best-fit parameters. 

\begin{figure}
   \centering
   \includegraphics[width=\columnwidth]{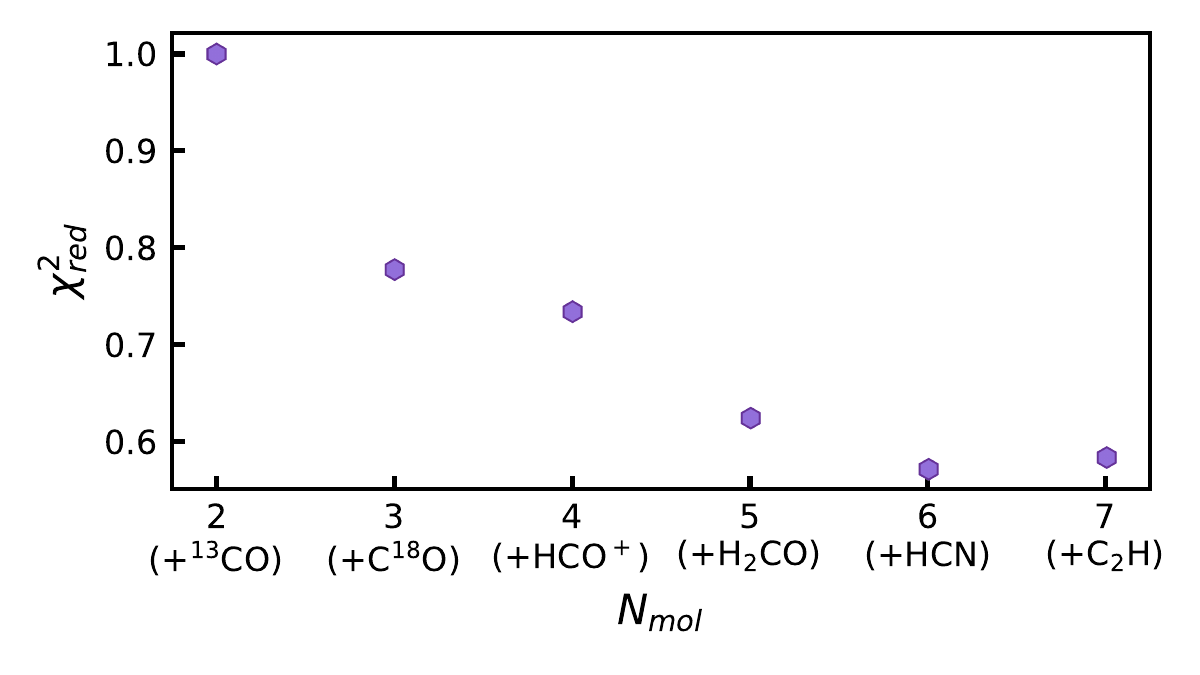}
      \caption{Measured $\chi^2_\text{red}$ (normalized to the maximum value) of the multi-molecule fits. We changed the number of molecules considered in the fit, adding the lines one by one.}
         \label{fig:chi2}
\end{figure}

To emphasize the importance of the multi-molecule approach, we fitted the rotation curves starting with two molecules and then progressively adding one molecule at a time. All the fits were performed with the same number of walkers, burn-in steps, and steps (200, 1000, 2000). To quantify the goodness of the fit, we show in Fig. \ref{fig:chi2} the $\chi^2_\text{red}$ obtained from all the fits, increasing one by one the number of fitted molecules. We normalized the $\chi^2_\text{red}$ to the maximum value, since the uncertainties on the extracted rotation curves are systematically underestimated (as explained above in this section).

As we can see from Fig. \ref{fig:chi2}, the most significant improvement is achieved with the inclusion of \ce{C^18O} in the fit, going from two to three molecules. This addition also induces the largest improvement in the dynamical estimates of the disk mass and scale radius: the values we obtain by fitting two and three molecules differ by $\lesssim 10\%$ and $\lesssim 6\%$ respectively for $M_\text{d}$ and $R_\mathrm{c}$. Moreover, the $\chi^2_\text{red}$ keeps decreasing with the further inclusion of lines in the multi-molecule fit. We added \ce{HCN and C2H} as last, since their rotation curves are the most affected by the beam smearing effect, as shown in Sect. \ref{subsec:beam_smearing}. We also found that the width of the posterior distribution shrinks as expected when fitting more rotation curves together, but our error bars on the velocity profiles are so small that they are still dominated by the systematics.

\begin{figure}
   \centering
   \includegraphics[width=\columnwidth]{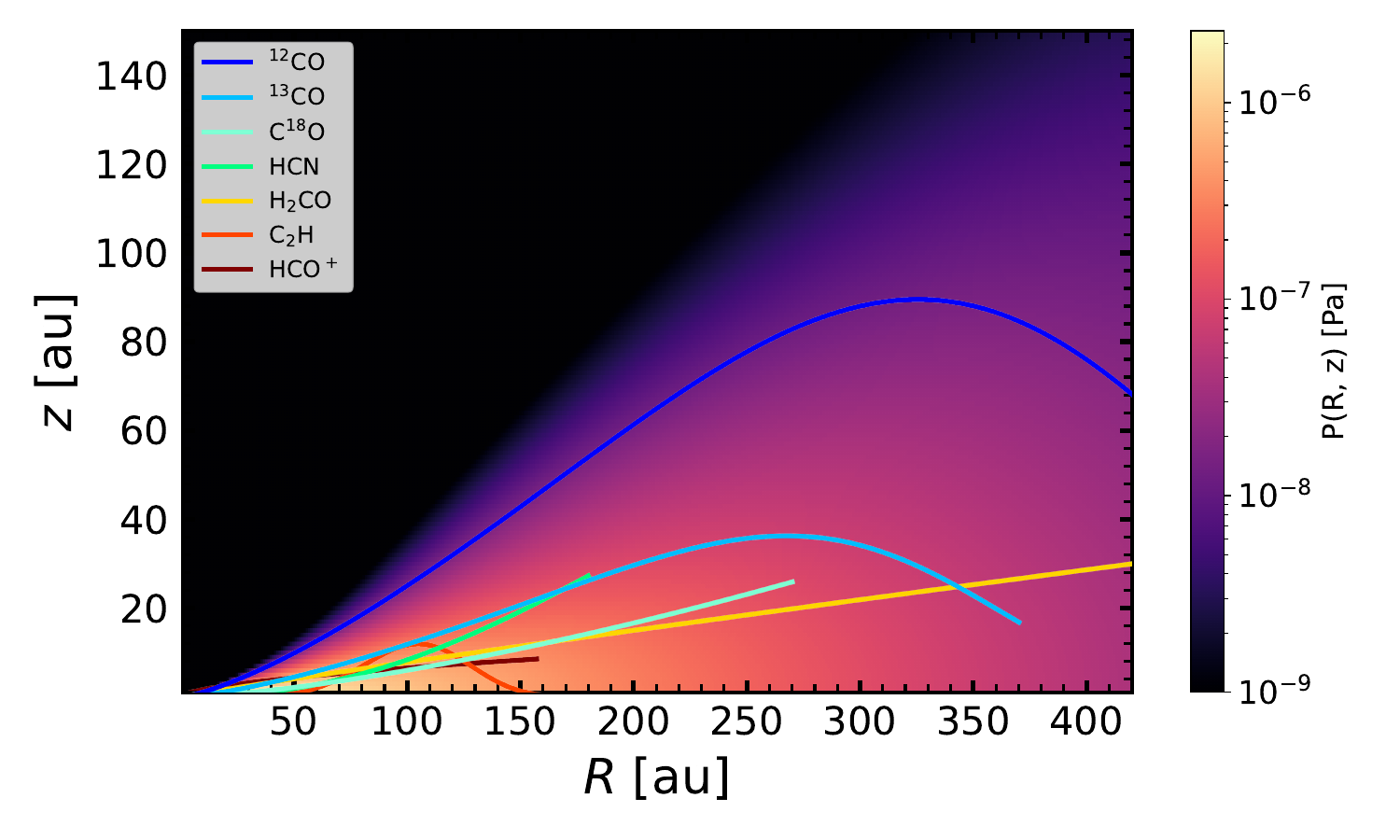}
      \caption{
      Retrieved 2D pressure structure $P(R,z)$ as a function of radius and height across the disk. The location of the selected molecular emitting layers are marked with colored solid lines, and highlight that we are tracing the pressure along an extended radial and vertical range in the disk.}
         \label{fig:pressure_plot}
\end{figure}

Knowing the disk mass and scale radius from the multi-molecule fit, and considering the thermal structure of the disk by \citet{maps4}, we can show the normalized 2D pressure structure of the disk in Fig. \ref{fig:pressure_plot}, following Eq. \ref{eq:pressure}. We note that the shape of the $P(R,z)$ profile is given by the assumed disk temperature $T(R,z)$, while the normalization term $P_\mathrm{mid}(R)$ is determined by the values for the disk mass and scale radius obtained by the fit of the rotation curves. We also show the location of the emitting layers of the tracers considered in the disk: the selected molecules probe the 2D pressure structure of the disk across its $(R,z)$ extent, from the upper surface down to the midplane.

\citet{martire} showed that thermal stratification plays an important role in explaining the discrepancies between \ce{^12CO} and \ce{^13CO} rotation curves. To do so, they compute the difference in the squared rotation velocities of the two molecules, with a proper normalization; this quantity strongly depends on the different heights of the tracers, and it is thus very useful to assess the relevance of thermal structure in disks. Hence, they show the expected behavior of this quantity according to both the isothermal and the stratified models, to see which one can better reproduce the observed data.
In this work, we perform the same calculation for all the considered molecules at different emitting heights, using our multi-molecule stratified model, to establish if thermal stratification has a strong impact on gas kinematics across the vertical extent of the disk. We compare squared rotation velocities of each molecule with the \ce{^12CO} one, and we illustrate the expected behavior of their normalized difference in the isothermal and in the vertically stratified scenarios. We show in Fig. \ref{fig:vel_diff_C18O} the results we obtained for \ce{C^18O}, while in Appendix \ref{app:vel_diff_others} we report the results for the other five molecules.

\begin{figure*}
   \centering
   \includegraphics[width=.8\textwidth]{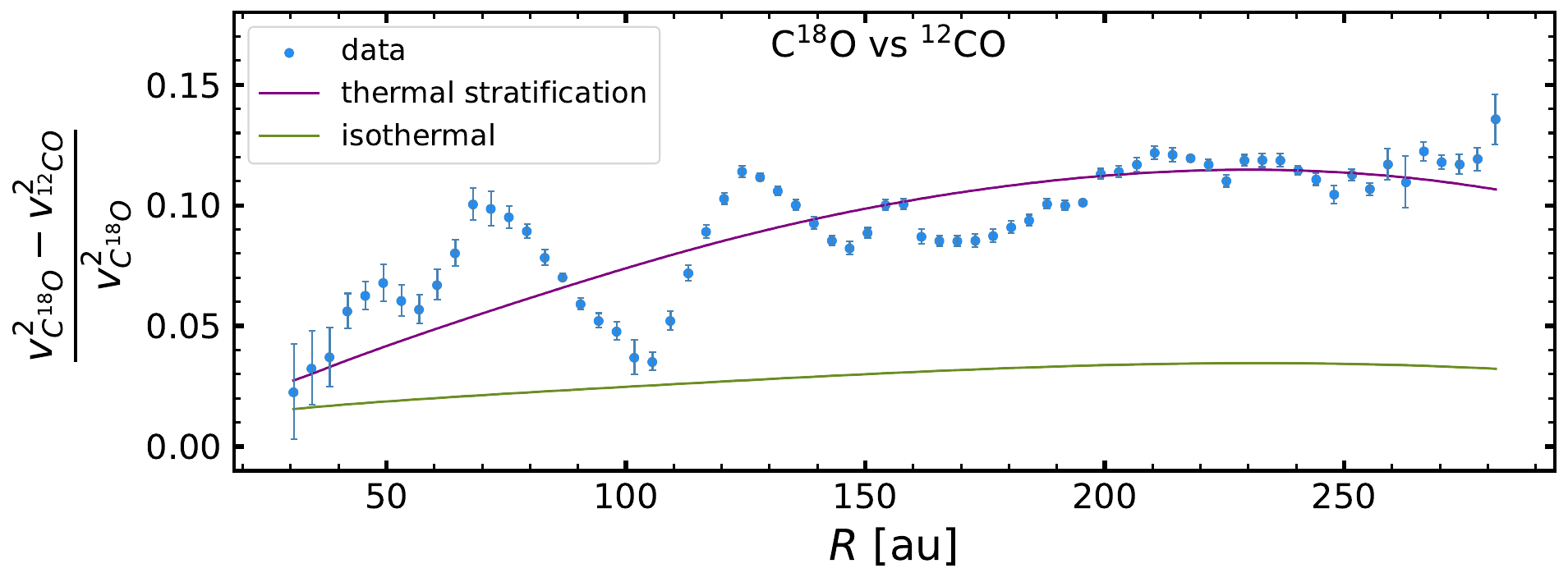}
      \caption{
      Normalized difference between the squared rotation velocities of \ce{C^18O} and \ce{^12CO}: Comparison between the data (light blue dots), the prediction from isothermal model (green line), and the prediction from thermally stratified model (purple line).}
         \label{fig:vel_diff_C18O}
\end{figure*}

From our results, it is clear that the thermally stratified prediction is more appropriate to describe the observed differences between the two rotation curves, while the isothermal model cannot reproduce the observed behavior. We observe that the effect of thermal stratification is important also in the lower layers located closer to the disk midplane, as shown in Figs. \ref{fig:vel_diff_C18O} and \ref{fig:vel_diff_others}. Temperature leaves a clear imprint in the kinematics of molecular lines originating from the whole vertical extent of the disk,  in the upper layers (as proved by \citealt{martire}) and in the layers close to the midplane. Once this is included in the model, this further large-scale modification to the pressure can be distinguished and measured efficiently.

Considering Fig. \ref{fig:vel_diff_C18O}, we point out that the difference in rotation velocity between the two isotopologs increases with radius and then starts decreasing at $\sim 200$ au; this trend is well reproduced by the stratified model, contrarily to the isothermal one. This discrepancy between the two models we considered for the rotation curves lies in the pressure gradient term, which includes an additional contribution depending on the height above the midplane when accounting for thermal vertical structure:
\begin{equation}
\begin{aligned}
    \frac{R}{\rho}\frac{dP}{dR}(R,z) 
    &= -v_\text{k}^2 \left[ \gamma^\prime + (2-\gamma)\left(\frac{R}{R_\mathrm{c}}\right)^{2-\gamma} - \right. \\
    &\quad\left. - \frac{\text{d}\log(fg)}{\text{d}\log R}\right]\left(\frac{H}{R}\right)_\text{mid}^2 f(R,z).
\end{aligned}
\end{equation}
In particular it can be verified that, for the thermal structure we assumed (i.e., the selected functions $f$ and $g$ in Eqs. \ref{eq:temp}, \ref{eq:cs}), this term remains negative across the whole radial extent of the disk and it decreases faster for \ce{^12CO} than for \ce{C^18O} as a function of radius, up to $\sim 200$ au. From this point onward, the pressure contribution keeps decreasing only for \ce{C^18O} while it remains approximately constant for \ce{^12CO}. Physically, this translates into a stronger effect of the pressure gradient in the higher layers with respect to the lower ones, when vertical thermal stratification is taken into account. For this reason, the gas rotation is slowed down at elevated heights more than near the midplane, because of the larger negative contribution of the pressure gradient. This behavior depends on the thermal structure of the disk and we only tested it for the temperature prescription we considered; nevertheless, we expect to observe the same result when describing the disk thermal structure with a similar function, both radially and vertically monotonic.
The difference between the predictions of the two models becomes less significant when the emitting layer of the considered molecule is closer to the \ce{^12CO} one: this is clear from the results for HCN shown in Appendix \ref{app:vel_diff_others}.

In Fig. \ref{fig:vel_diff_C18O} we also see an oscillating behavior of the difference of the velocities squared. These modulations are likely correlated to the radial pressure substructure exhibiting peaks and troughs, as imprinted in molecular rotation curves \citep{teague2019, izquierdo2023}. 
As we are not accounting for pressure modulated substructures in our model, we are not able to recover the oscillating behavior of the measured velocity difference. An important improvement would be to include these small-scale pressure oscillations in order to perform a more exhaustive fit of the curves. It is worth highlighting that these additional pressure effects are seen in the difference of the velocities squared, extracted from two different emission heights, which thus suggests different amplitudes of the pressure substructure at distinct heights in the disk. This indicates that rotation curves can be a powerful tool to characterize the 3D structure of gaps, which should be further investigated in the future.

In summary, we can use multi-molecule emission to investigate and directly probe the 2D pressure structure of disks, one of the main physical properties that shape disks dynamics; we can measure the pressure effects on a global scale (i.e., extended pressure gradients along the radial and vertical range) and on a local scale (i.e., localized pressure-modulated substructures).

\subsection{Dependence on the emitting layers}
\label{subsec:layers_dependency}
In this section we analyze how changing the considered molecular emitting layers can impact our results on the dynamical estimates of $M_\star$, $M_\text{d}$, and $R_\text{c}$.

As a first test, we repeated the whole identical routine of rotation curves extraction and multi-molecule fit of the curves, considering flat emission for all the molecules. We note that this scenario is extreme and does not well represent the real system, as we are assuming that all the molecules emit from the midplane, which is clearly not realistic as \ce{^12CO, ^13CO, and HCN} are known to emit from very elevated layers. In this case, we obtain the following best-fit estimates: $M_\star=1.7827\pm 0.0003$ M$_\odot$, $M_\text{d}=0.1396\pm 0.0003$ M$_\odot$, and $R_\text{c}=174.8\pm 0.3$ au. Thus, the differences in the final dynamical estimates of the parameters are within 10\% for M$_\star$, 18\% for $M_\text{d}$, and 20\% for $R_\text{c}$.
We note that these differences are mainly driven by the overestimation of the rotation curve of \ce{^12CO} in the outer regions of the disk; in fact, we fit a lower value for M$_\star$ and a higher value for $M_\text{d}$, which is strongly related to the disk outer regions. However, in this case we are assuming \ce{^12CO} is emitting from the midplane and we know that this scenario is unrealistic. However, a much more realistic scenario where we consider flat emission only for lower S/N lines and/or molecules emitting close to the midplane (\ce{HCO^+,\ H2CO,\ C2H, and C^18O}) would impact the results far less, as the extracted curves in the flat and nonflat scenarios are compatible (as explained in Sect. \ref{subsec:curves_extraction} and shown in Fig. \ref{fig:flat_emission_diff}).

As a second test, we repeated the whole analysis adopting nonparametric emitting layers for all the molecules, as retrieved by \citet{paneque}. In Appendix \ref{app:nonpar_layers} we show the corner plot obtained from the multi-molecule fit, with the dynamical estimates of the stellar mass, disk mass, and scale radius. The best-fit values we obtain from the fit in the nonparametric case differ from the parametric case results by $\lesssim 1\%$ for $M_\star$, $\lesssim 30\%$ for $M_\text{d}$, and $\lesssim 13\%$ for $R_\text{c}$. We see that the shape of the emitting surface has an impact mainly on the disk mass, that is affected by the systematics described by \citet{veronesi2024} and \citet{andrews2024}. The difference we measure for the disk mass estimate is in agreement with the results by \citet{veronesi2024} and \citet{andrews2024}, predicting an uncertainty close to $\sim 25 \%$.
Also, in Appendix \ref{app:nonpar_layers} we show the same plot as Fig. \ref{fig:vel_diff_C18O} for the nonparametric case, in which we recover exactly the same trends as the parametric case; thus, the same conclusions we drew in Sect. \ref{subsec:multimol_fit} also apply to the nonparametric scenario.


\section{Conclusions}
\label{sec:5_discussion_conclusions}

In this work we explored the potential of a multi-molecule approach to gas kinematics to accurately characterize the 2D ($R,z$) pressure structure of disks. We focused our analysis on the HD~163296 protoplanetary disk, and investigated the 2D structure by analyzing seven molecular tracers originating at different heights across the disk vertical extent to gain insights into the physical properties across the distinct disk layers.
Considering the disk 2D temperature structure as estimated through the optically thick emission layering by \citet{maps4} and the emission heights by \citet{paneque}, we extended existent tools to extract the rotation curves along these layers. We then simultaneously fitted all the curves for the first time with a multi-molecule ($\geq 2$) model of a disk including thermal stratification. Our results show the following:

\begin{itemize}
    \item[$\bullet$] The 2D pressure structure leaves a measurable kinematical imprint in the gas rotation curves that is important across the whole disk vertical extent, also close to the midplane. Therefore, thermal stratification must be included in models to correctly reproduce the measured velocity differences at distinct heights, extending the approach by \citet{martire} closer to the midplane.
    \item[$\bullet$] There are pressure substructures linked to the existence of dust gaps (not considered in our model) that can be seen from the rotation curves analysis and clearly exhibit a vertical dependence.
    \item[$\bullet$] A multi-molecule fit ($\geq 2$) of the rotation curves including the contribution from vertical stratification can be performed efficiently and dynamical estimates can be constrained more accurately (within the limits of our parametric method) for the stellar mass ($M_\star=1.89$ M$_\odot$), disk mass ($M_\text{d}=0.12$ M$_\odot$), and disk scale radius ($ R_\text{c}=143$ au), in agreement with \citet{martire}.
    \item[$\bullet$] The main results we obtain do not depend significantly on the chosen parameterization for the molecular emitting layers. The disk mass is the parameter that is more affected by these systematics ($\lesssim 30\%$).
    \item[$\bullet$] We can retrieve the 2D pressure structure of the disk by knowing the 2D temperature profile, the disk mass, and the scale radius from the rotation curves fit.
    \item[$\bullet$] Spectra from molecular lines with hyperfine splitting can be efficiently leveraged by extending standard methodologies to enhance precision and accuracy in the retrieved rotation curves.
    \item[$\bullet$] Optical depth profiles can be easily constrained knowing the 2D temperature and the molecular emitting layers only, with reasonable accuracy. These results are comparable to estimates from other independent and more complicated methods.
\end{itemize}

A first interesting development of this work would be to extract the temperature field directly from the rotation curves instead of assuming it from \citet{law2021}, by fitting for the temperature structure as imprinted in the vertically separated rotation curves. With this method it would be possible to characterize the 2D temperature field $T(r,z)$ directly from molecular rotation curves without the restriction to optically thick emitters. We leave this analysis to subsequent work.

We also emphasize that the simultaneous characterization of the 2D pressure structure and temperature scalar field has the potential to reveal the 2D gas density structure in disks. A multi-molecule kinematical analysis could then represent a solid further step toward a more comprehensive physical and chemical view of the 2D structure of disks.

\begin{acknowledgements}
This paper makes use of the following ALMA data: ADS/JAO.ALMA\#2018.1.01055.L ALMA is a partnership of ESO (representing its member states), NSF (USA) and NINS (Japan), together with NRC (Canada), MOST and ASIAA (Taiwan), and KASI (Republic of Korea), in cooperation with the Republic of Chile. The Joint ALMA Observatory is operated by ESO, AUI/NRAO and NAOJ. V.P. and S.F. are funded by the European Union (ERC, UNVEIL, 101076613). Views and opinions expressed are however those of the author(s) only and do not necessarily reflect those of the European Union or the European Research Council. Neither the European Union nor the granting authority can be held responsible for them. S.F. acknowledges financial contribution from PRIN-MUR 2022YP5ACE. CL has been supported by the UK Science and Technology research Council (STFC) via the consolidated grant ST/W000997/1. GL acknowledges the European Union’s Horizon 2020 research and innovation program under the Marie Sklodowska-Curie Grant Agreement No 823823 (DUSTBUSTERS). The authors thank the anonymous referee for valuable comments that improved the scientific value of this work.

\end{acknowledgements}
\bibliographystyle{aa}
\bibliography{bibliography}

\begin{appendix}

\section{Flat emission scenario}

\begin{figure}[h!tbp]
    \centering
    \includegraphics[width=.8\columnwidth]{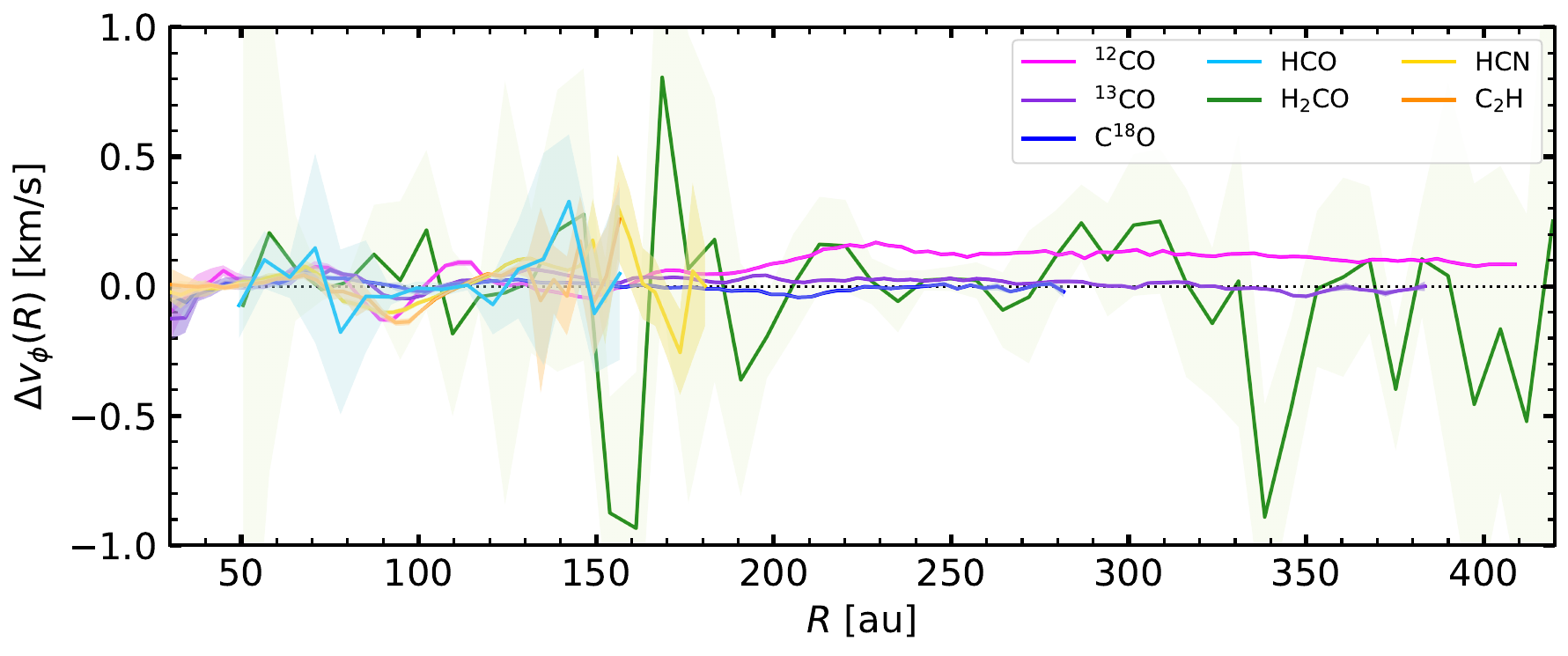}

    \caption{Difference between the rotation curves assuming flat and elevated emission, for each of the considered lines.}

    \label{fig:flat_emission_diff}
\end{figure}

For every considered line, in Fig. \ref{fig:flat_emission_diff} we show the difference between the extracted rotation curves in case of flat and elevated emission, where the error on each point is evaluated as the sum in quadrature of the errors on the two curves. The errors are generally small, as the uncertainties on the curves are in the first place underestimated, as explained in Sect. \ref{subsec:multimol_fit}. 
The largest discrepancies not compatible within the error bars are obtained for \ce{^12CO}, \ce{^13CO}, and HCN, for which the flat assumption is clearly incorrect, as these molecules emit from elevated layers in the disk. This was expected, and these molecules do not represent an issue as their emission has a high S/N and their emitting layer is well-constrained. As expected, we obtain a large discrepancy for \ce{^12CO}, especially at larger radii: in case of flat emission, the curve is overestimated in the outer regions of the disk ($R>200$ au), leading in turn to an overestimate of the disk mass, as described in Sect. \ref{subsec:layers_dependency}.

\FloatBarrier

\section{Sample extension: Beam smearing effect}
\label{app:HCN}
\begin{figure}[h!tbp]
    \centering
    \includegraphics[width=.9\columnwidth]{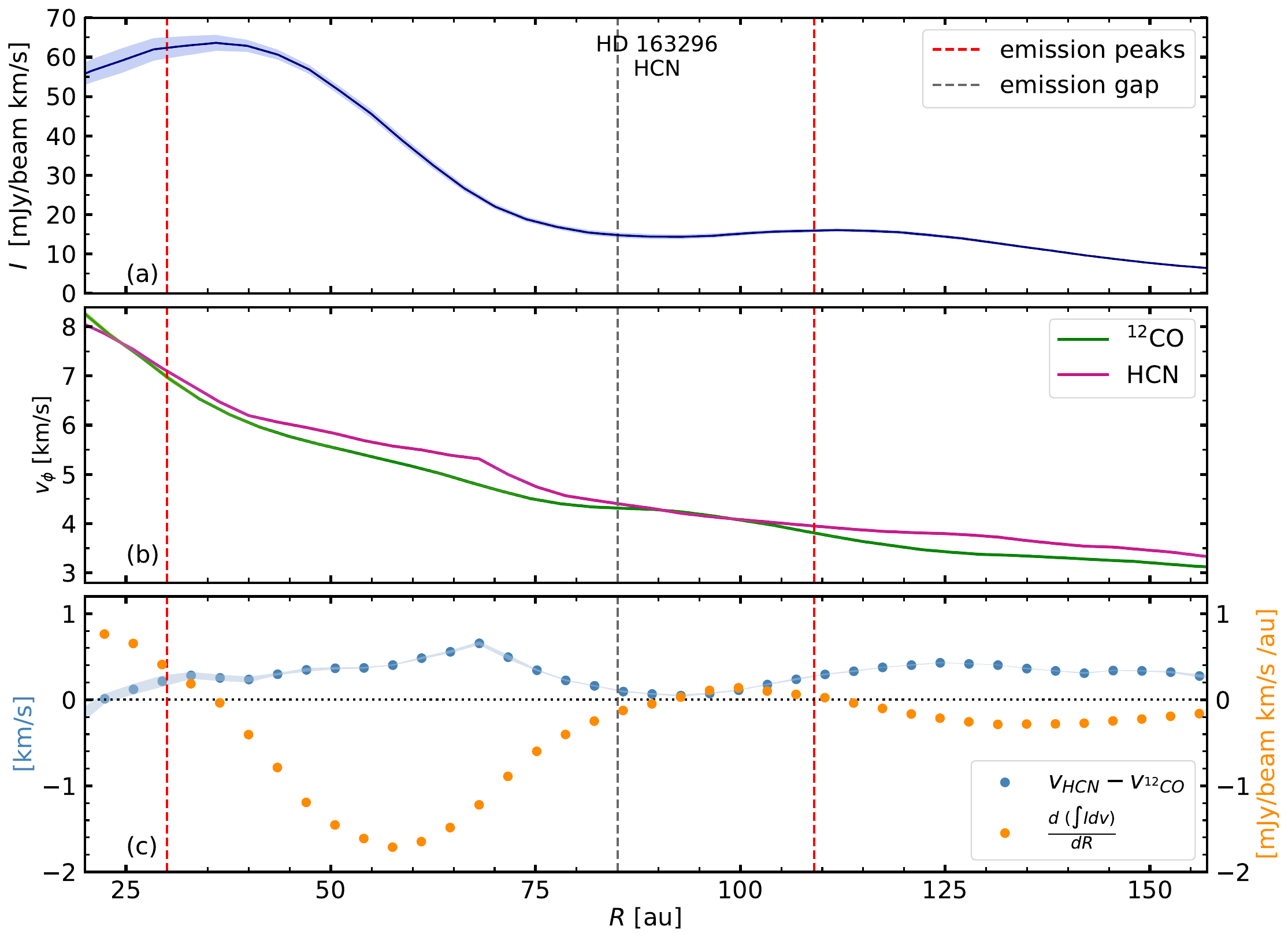}

    \caption{
    Beam smearing effect for HCN.\\
   \textit{Panel (a)}: Azimuthally averaged integrated intensity profile of the HCN line. Emission peaks are marked with dashed red lines and the emission gap is highlighted with a dashed grey line. \textit{Panel (b)}: Rotation curve of HCN (purple line) compared to the one of $^{12}$CO (green line). The rotation velocity of HCN is super-Keplerian just before the gap and sub-Keplerian after. It shows the opposite behavior in correspondence of the emission peaks. \textit{Panel (c)}: Anticorrelation between the HCN integrated intensity gradient and its difference in rotation velocity compared to $^{12}$CO. The two quantities show an opposite modulation and they cross at the radii corresponding to the emission gap or peaks.}
    \label{fig:anticorr_HCN}
\end{figure}

We show in Fig. \ref{fig:anticorr_HCN} the beam smearing effects for HCN, as previously shown in Fig. \ref{fig:anticorr} for \ce{C2H} in Sect. \ref{subsec:beam_smearing}.

\FloatBarrier

\section{\ce{H2CO}: A complicated case}
\label{app:H2CO}

\begin{figure}[h!]
   
   \includegraphics[width=\columnwidth]{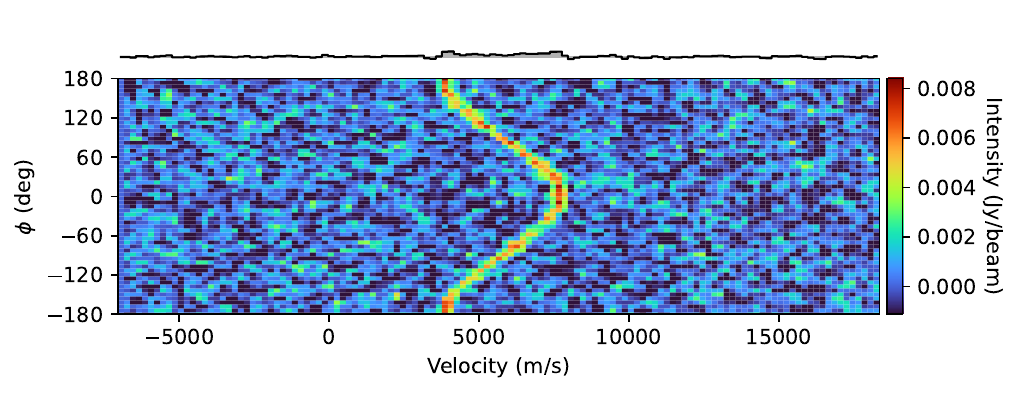}
   \includegraphics[width=\columnwidth]{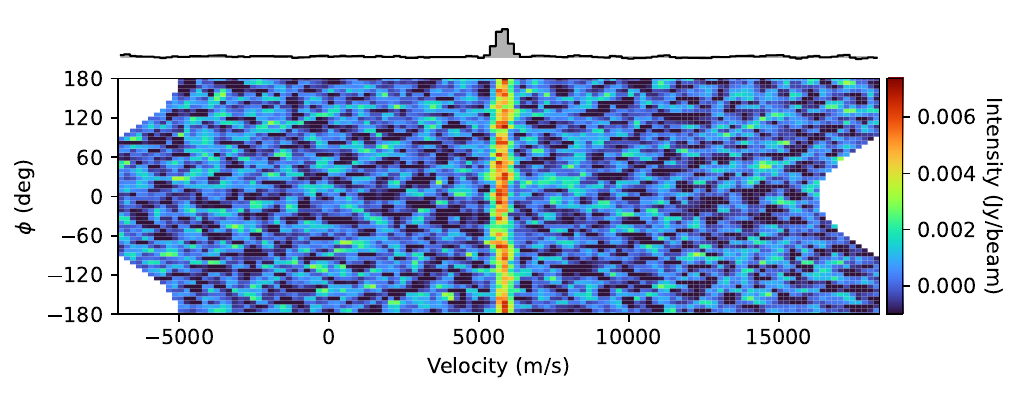}
   \includegraphics[width=\columnwidth]{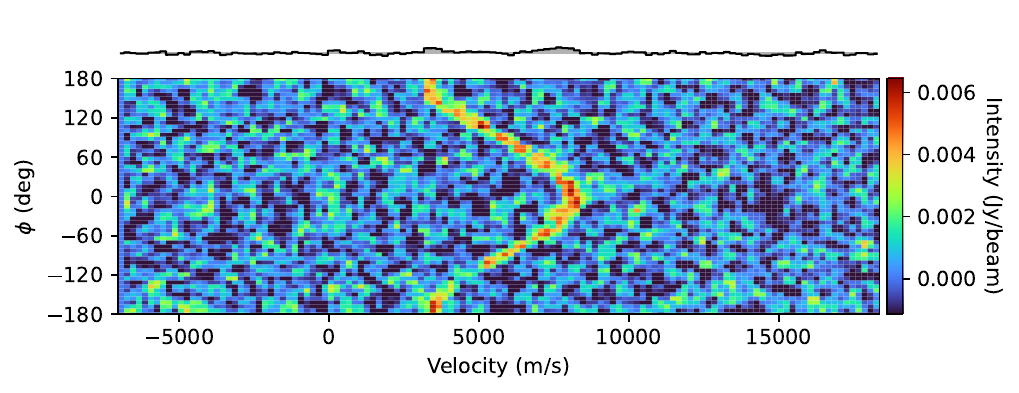}
   \includegraphics[width=\columnwidth]{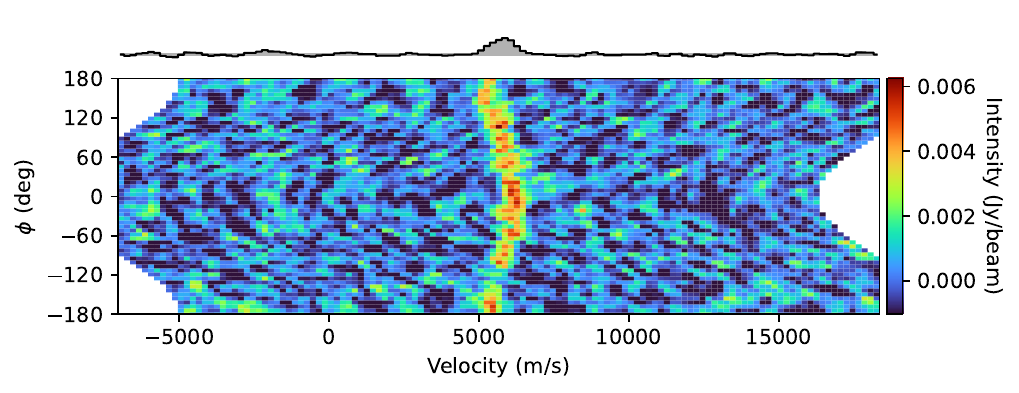}
   
      \caption{River plots before and after the alignment procedure with the SHO method from \eddyfont, at $R\sim 240$ au (two upper panels) and $R\sim 150$ au (two bottom panels).}
         \label{fig:river_plots_H2CO}
 \end{figure}

As mentioned in Sect. \ref{subsec:multimol_fit}, the rotation curve extraction with \eddyfont \ does not provide reliable results for a small range of radii around $R\sim 150$ au, as the signal-to-noise in correspondence of this emission gap is too low to allow an efficient reconstruction of the profile. We note that the extraction in this region is not reliable by looking at the river plots at the corresponding radii and at how efficiently they are aligned. We show in Fig. \ref{fig:river_plots_H2CO} the river plots, before and after the alignment with the SHO method, at two different radii: $R\sim 240$ au and $R\sim 160$ au. We see that in the first case the alignment procedure works perfectly and the aligned spectra form a straight line, while in the second case (corresponding to the emission gap radius) the method fails in aligning the intensity peaks.

\FloatBarrier

\section{Thermally stratified fit: Corner plot}
\begin{figure}
   \centering
   \includegraphics[width=\columnwidth]{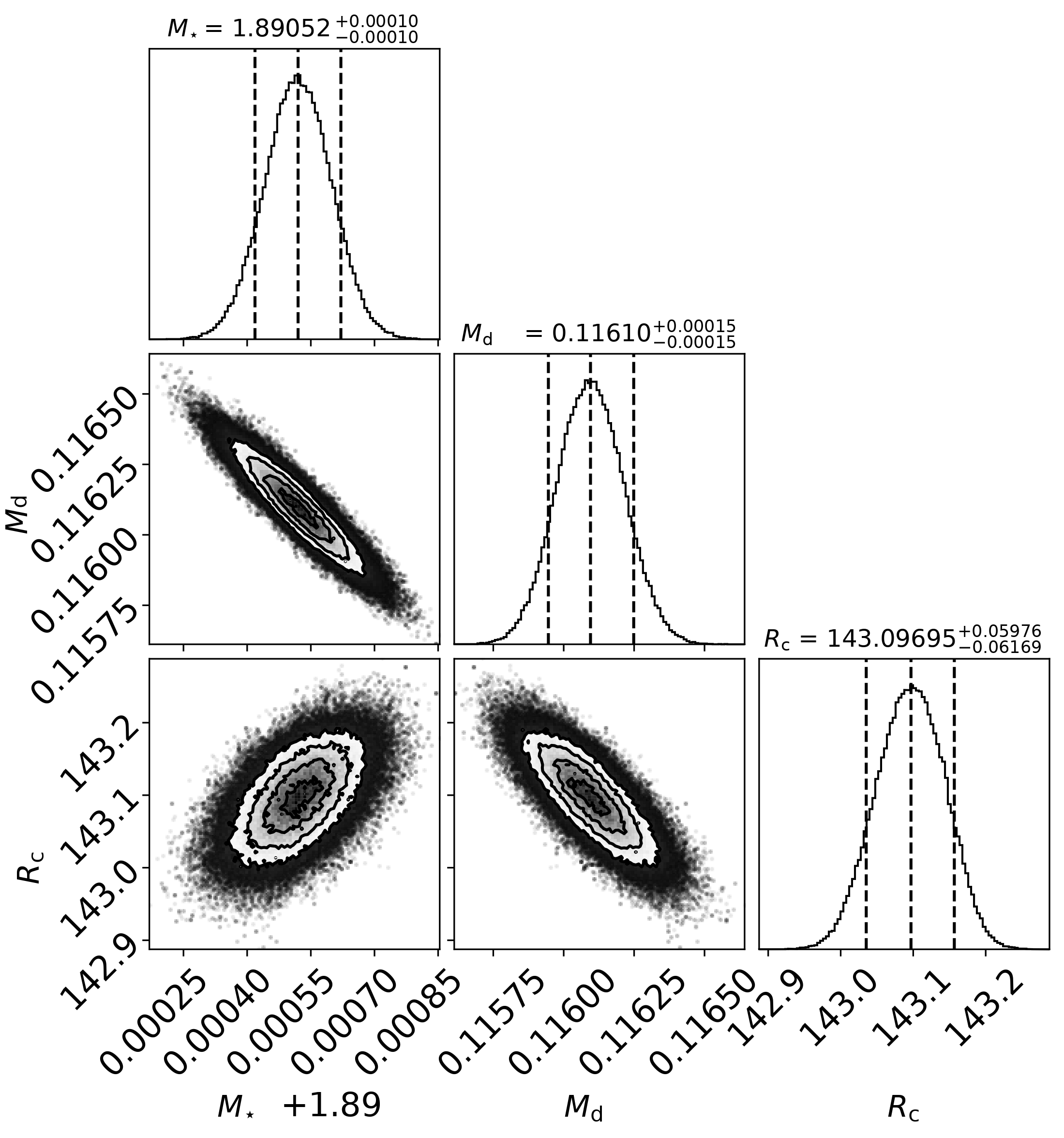}
   
      \caption{Corner plot of the multi-molecule stratified fit performed with DySc on the seven considered rotation curves. The uncertainties reported in the plot represent the default errors we obtain from the MCMC process, not the one we consider for our analysis, as explained in Sect. \ref{subsec:multimol_fit}.}
         \label{fig:cornerplot}
\end{figure}

We show in Fig. \ref{fig:cornerplot} the corner plot for the simultaneous multi-molecule fit performed with DySc on our seven extracted rotation curves.
\label{app:cornerplot}
\FloatBarrier
\section{Nonparametric emitting layers}
\label{app:nonpar_layers}

\begin{figure}[h!]
   \centering
   \includegraphics[width=\columnwidth]{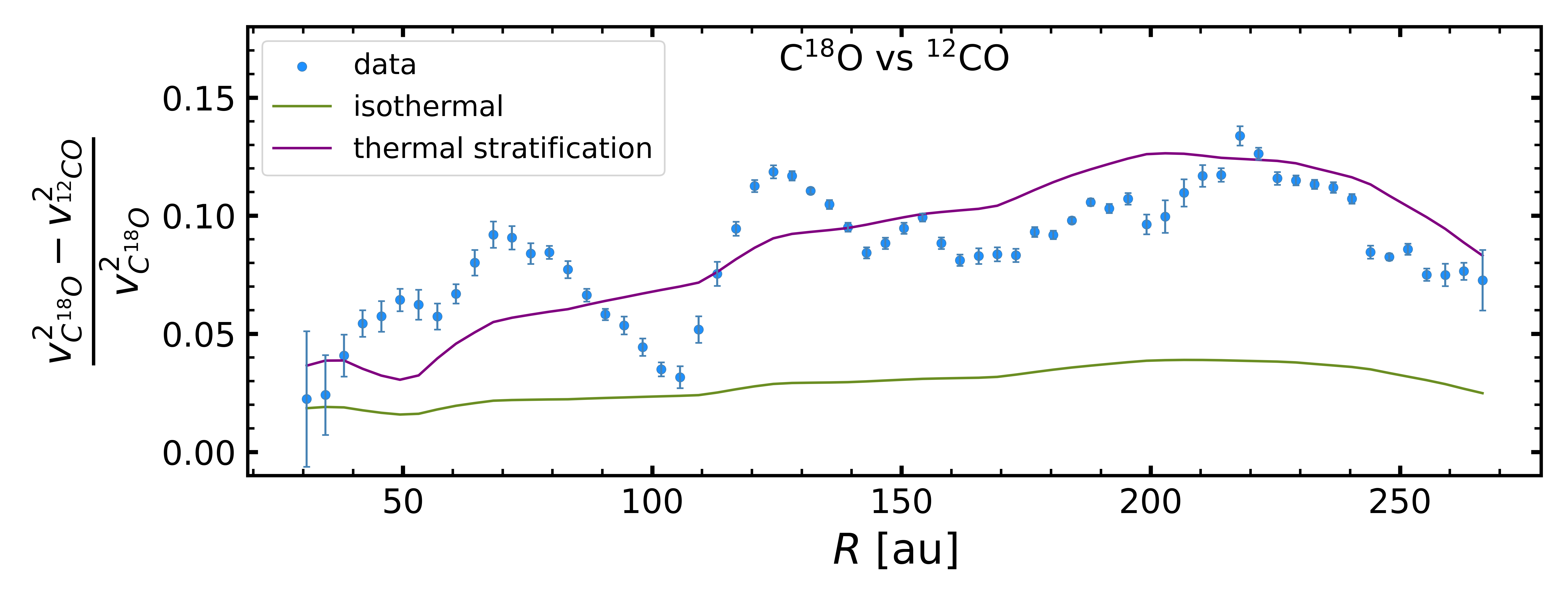}
      \caption{Normalized difference between the squared rotation velocities of \ce{C^18O} and \ce{^12CO}, assuming nonparametric emitting layers for all the molecules: comparison between the data (light blue dots), the prediction from isothermal model (green line) and from thermally stratified model (purple line).}
         \label{fig:diff_C18O_nonpar}
\end{figure}

\begin{figure}
   \centering
   \includegraphics[width=\columnwidth]{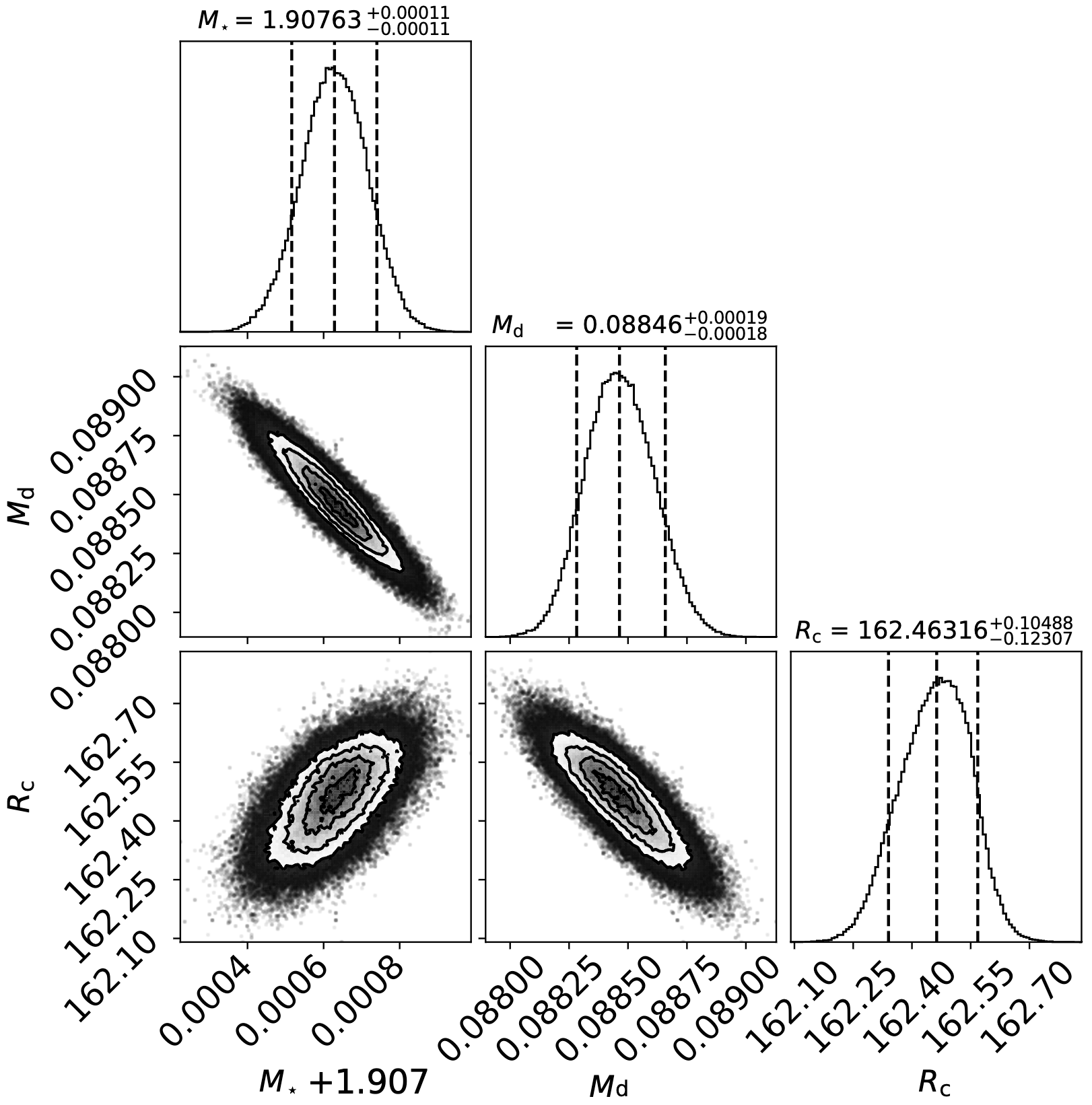}
      \caption{Corner plot of the multi-molecule stratified fit performed with DySc on the seven considered rotation curves, assuming nonparametric emitting layers for all the molecules.}
         \label{fig:corner_7mol_nonpar}
\end{figure}

We show in Fig. \ref{fig:corner_7mol_nonpar} the corner plot obtained from the multi-molecule fit assuming nonparametric emitting layers for all the molecules, with the dynamical estimates of the stellar mass, disk mass, and scale radius.
We also show in Fig. \ref{fig:diff_C18O_nonpar} the same plot as Fig. \ref{fig:vel_diff_C18O} for the nonparametric case.

\FloatBarrier

\section{Complete sample: Thermally stratified versus isothermal model}
\label{app:vel_diff_others}

\begin{figure}
   \centering
   \includegraphics[width=\columnwidth]{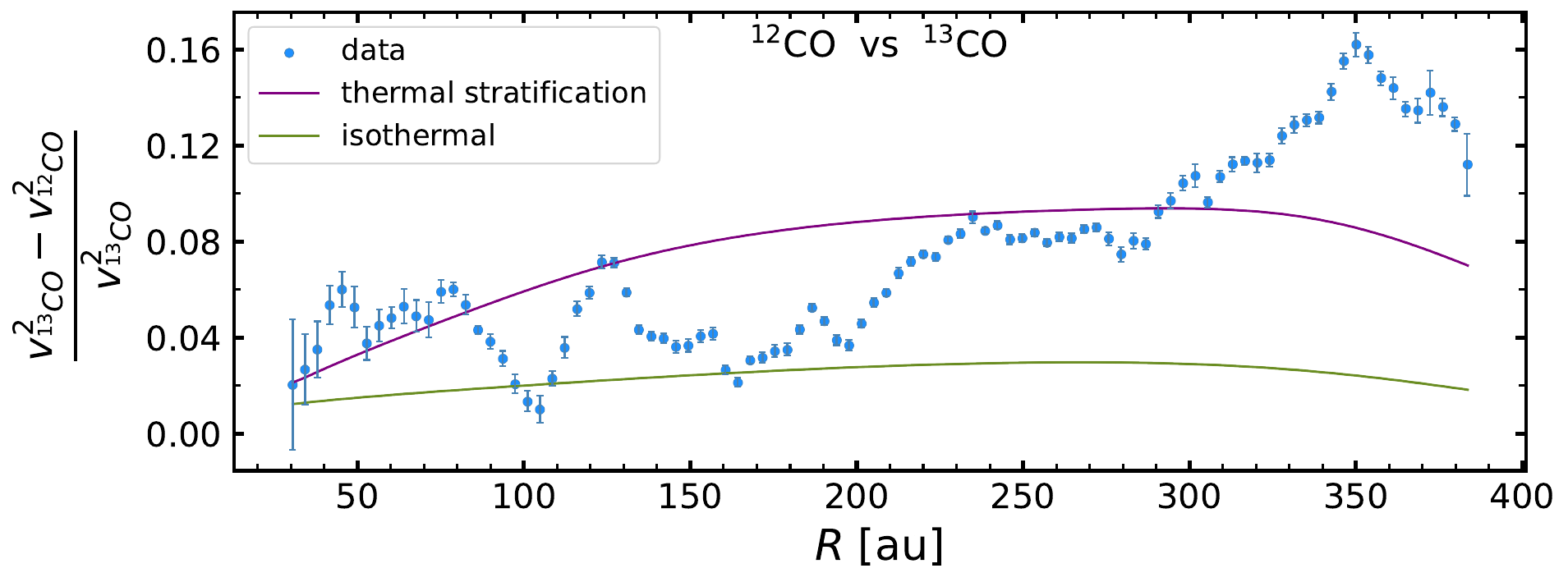}
   \includegraphics[width=\columnwidth]{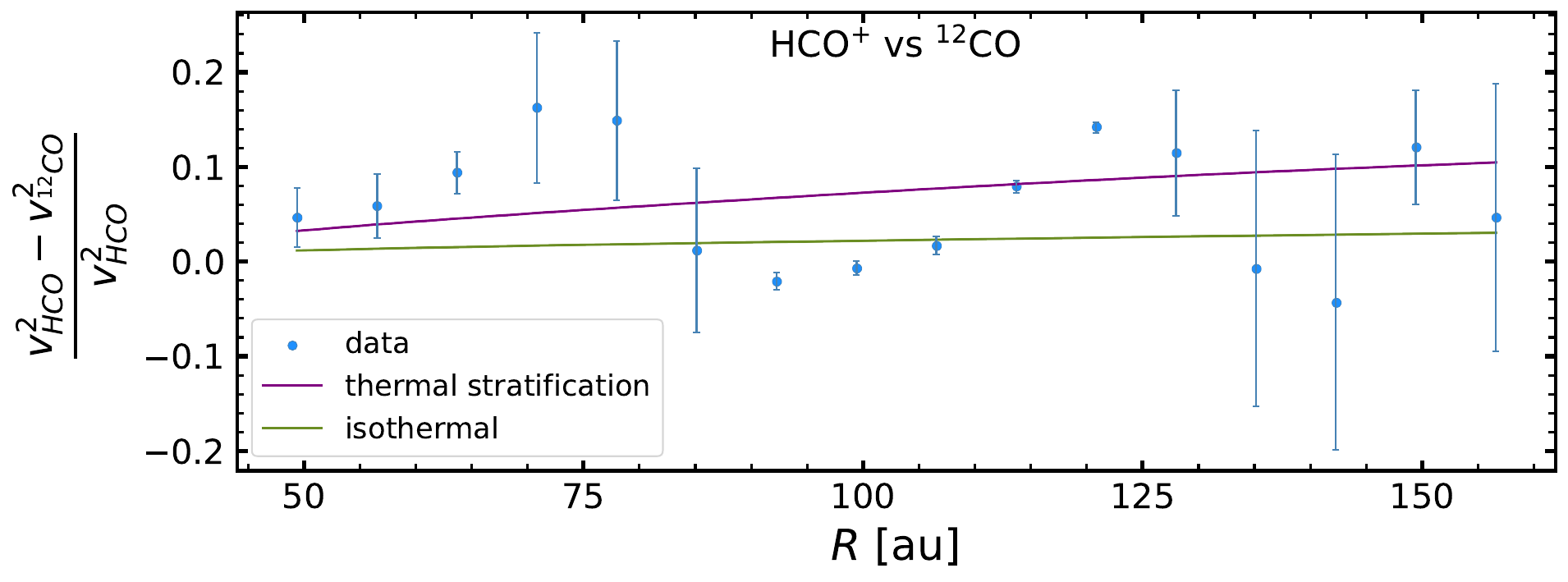}
   \includegraphics[width=\columnwidth]{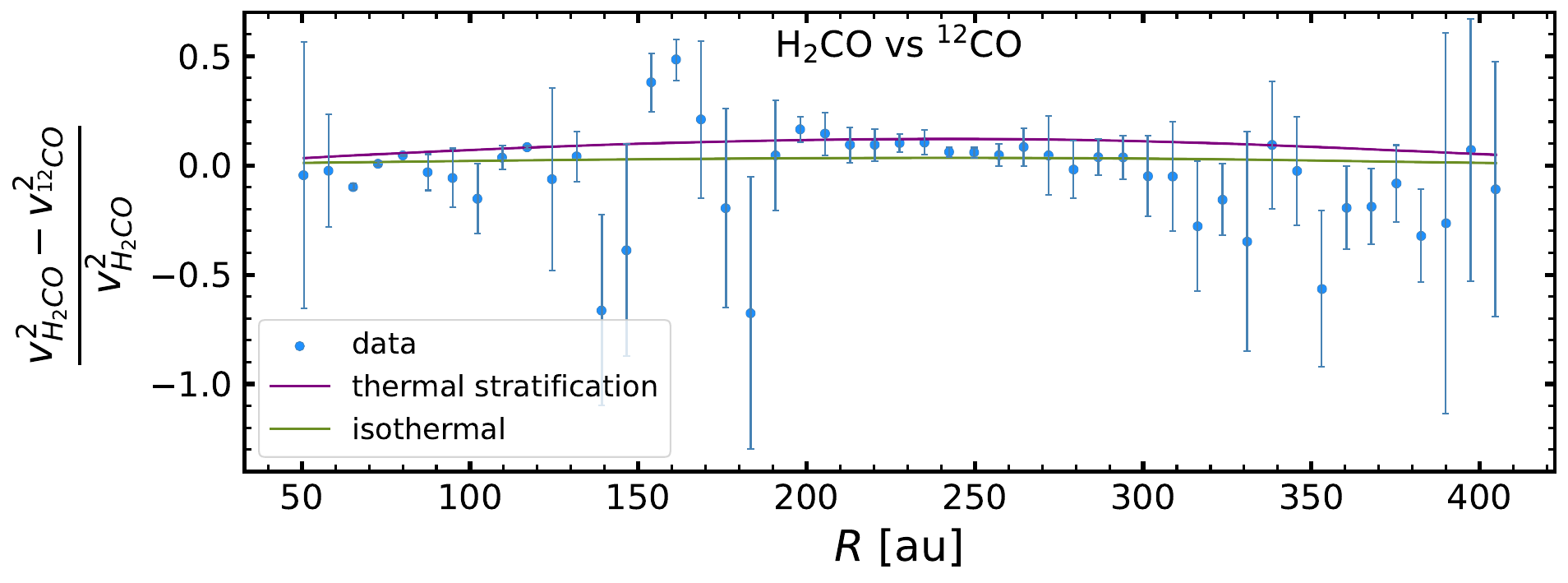}
   \includegraphics[width=\columnwidth]{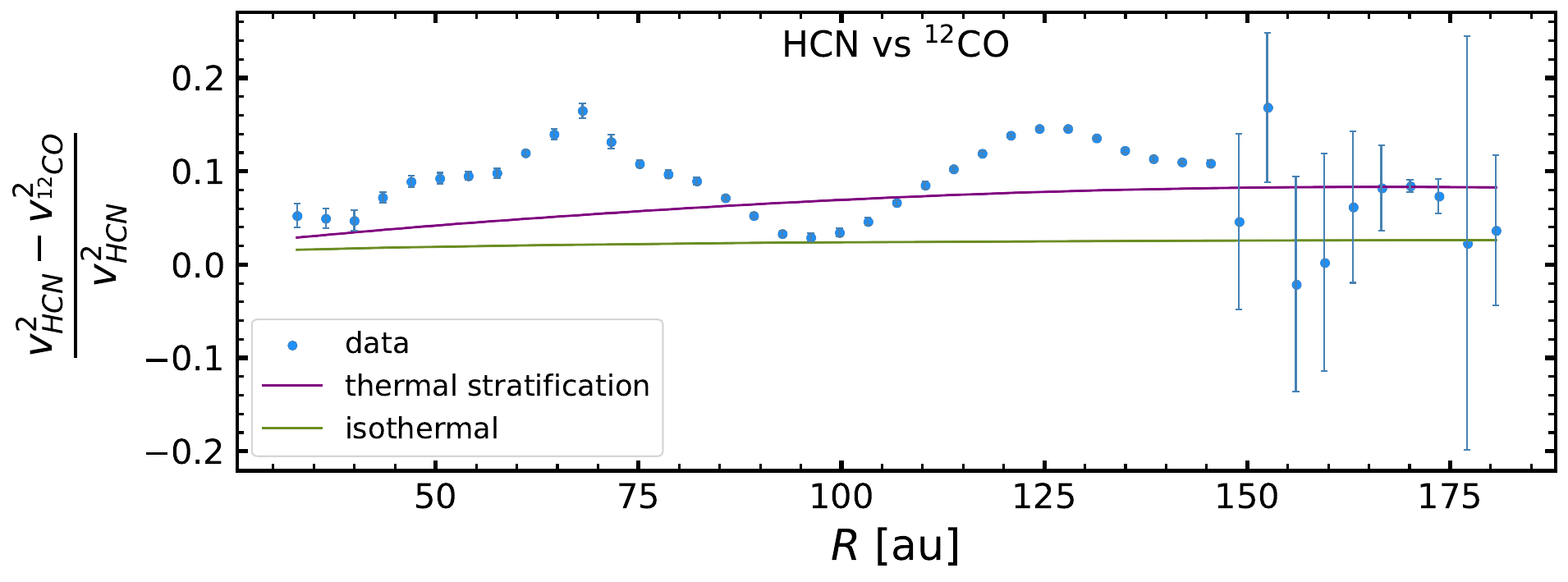}
   \includegraphics[width=\columnwidth]{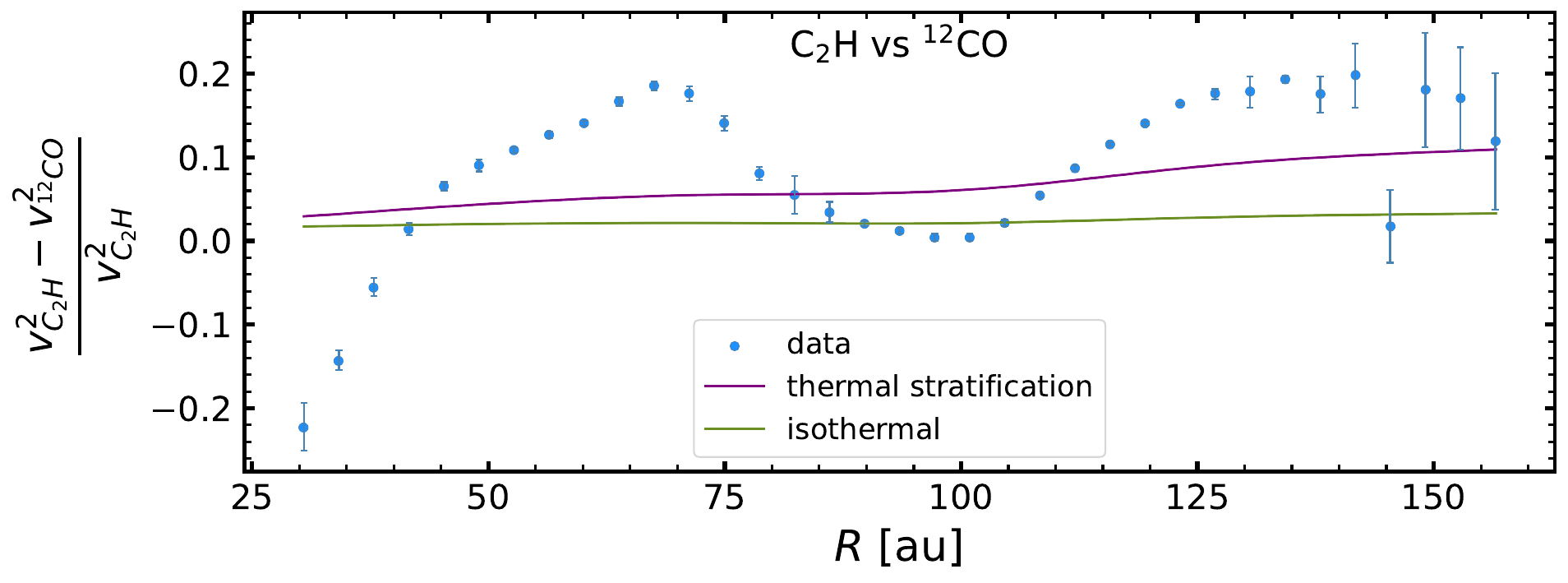}
      \caption{Normalized difference between the squared rotation velocities of each molecule and \ce{^12CO}: comparison between the data (light blue dots), the prediction from isothermal model (green line) and from thermally stratified model (purple line).}
         \label{fig:vel_diff_others}
\end{figure}

Here we show in Fig. \ref{fig:vel_diff_others} the normalized difference between the squared rotation velocities of respectively \ce{^13CO}, \ce{HCO^+}, \ce{H2CO}, HCN, and \ce{C2H} with respect to the \ce{^12CO}. 
We do not discuss in detail \ce{H2CO}, since the rotation curve extraction proved to be quite difficult and we do not have the necessary accuracy to draw a reasonable conclusion.
We also do not recover any significant difference between the two considered models for what concerns \ce{HCO^+}.

On the contrary, we find a significant improvement in the fit for \ce{^13CO}, when considering thermal structure: in this case, as for \ce{C^18O}, the stratified model better reproduces the measured velocity difference. This is the case also for HCN and \ce{C2H}, even if the difference between the two models is less appreciable here because in the considered range of radii their emitting layers are closer to the \ce{^12CO} with respect to the \ce{^13CO}. Moreover, for these two molecules we can clearly see a global oscillation in the observations that can be traced back to the distortion of the rotation curves due to the beam smearing effect, as explained in Sect. \ref{subsec:beam_smearing}. Finally, for \ce{^13CO} we find small-scale pressure modulated oscillations similar to what we see for \ce{C^18O}.

We observe that, especially in the \ce{^13CO} case, for some radial ranges ($130-210$ au, $>330$ au) even the stratified model struggles to reproduce the data. This can be traced back to the inflexibility of the model in this case, caused by the not complete suitability of the temperature parameterization from \citet{dullemond2020} we are assuming, as the thermal structure might be more complex. 

\end{appendix}
\end{document}